\documentclass[12pt]{article}

\newcommand{\sla}[1]{#1\kern-0.55em/}
\usepackage{graphicx}
\usepackage{dcolumn}
\usepackage{bm}
\usepackage{cite}
\usepackage{amsmath}
\usepackage{amssymb}
\usepackage{color}

\begin{document}
\begin{flushright}
CERN-PH-TH-2011-052\\
LYCEN 2011-01
\end{flushright}
\begin{center}
\Large\bf\boldmath
\vspace*{0.8cm} Anomaly mediated SUSY breaking scenarios in the light of cosmology and
in the dark (matter)
\unboldmath
\end{center}

\vspace{0.6cm}
\begin{center}
A. Arbey$^{1,2,3,}$\footnote{\tt alexandre.arbey@ens-lyon.fr}, A. Deandrea$^{1,}$\footnote{\tt deandrea@ipnl.in2p3.fr} and A. Tarhini$^{1,}$\footnote{ \tt tarhini@ipnl.in2p3.fr}\\[0.4cm]
{\sl $^1$ Universit\'e de Lyon, France; Universit\'e Lyon 1, CNRS/IN2P3, UMR5822 IPNL, F-69622~Villeurbanne Cedex, France}\\
\vspace{0.6cm}
{\sl $^2$ CERN Theory Division, Physics Department, CH-1211 Geneva 23, Switzerland}\\
\vspace{0.6cm}
{\sl $^3$ Centre de Recherche Astrophysique de Lyon, Observatoire de Lyon, \\
9 avenue Charles Andr\'e, Saint-Genis Laval cedex, F-69561, France; CNRS, UMR 5574; \\
Ecole Normale Sup\'erieure de Lyon, Lyon, France.}\\
\vspace{0.6cm}
\end{center}
\vspace{0.4cm}
\begin{abstract}
Anomaly mediation is a popular and well motivated supersymmetry breaking scenario. Different possible detailed 
realisations of this set-up are studied and actively searched for at colliders. Apart from limits coming from flavour, low energy 
physics and direct collider searches, these models are usually constrained by the requirement of reproducing the observations on 
dark matter density in the universe. We reanalyse these bounds and in particular we focus on the dark matter bounds both 
considering the standard cosmological model and alternative cosmological scenarios. These scenarios do not change the 
observable cosmology but relic dark matter density bounds strongly depend on them. We consider few benchmark points 
excluded by standard cosmology dark matter bounds and suggest that loosening the dark matter constraints is necessary 
in order to avoid a too strong (cosmological) model dependence in the limits that are obtained for these models. We also discuss
briefly the implications for phenomenology and in particular at the Large Hadron Collider.
\end{abstract}
%
\vspace{1.2cm}

\section{Introduction}
The search for supersymmetry and its breaking, in addition to the direct searches at LEP, B-factories, Tevatron and the 
Large Hadron Collider (LHC), is actively pursued using the WMAP limits on the relic density constraints. 
However the sensitivity of the lightest 
supersymmetric particle relic density calculation to the variation of the cosmological expansion rate before Big-Bang 
Nucleosynthesis (BBN), even if modest and with no consequences on the cosmological observations, can modify considerably 
the relic density, and therefore change the constraints on the supersymmetric parameter space \cite{Arbey:2008kv,Arbey:2009gt}. 
In the standard cosmology the dominant component before BBN is radiation, however energy density and entropy content can be 
modified. In the following we 
consider the impact of different scenarios of alternative cosmologies. The precision of the WMAP data should therefore not make 
us forget the hypothesis which are implied by the use of standard cosmology. We discuss in the following the implications of 
precision B-physics, direct searches and cold dark matter relic abundance for the case of anomaly mediated models, from a 
minimal anomaly mediation supersymmetry breaking \cite{amsb}, to mixed moduli-anomaly 
mediated \cite{Choi:2005uz} and to hypercharge anomaly mediation \cite{Dermisek:2007qi}. We also discuss similar 
supersymmetry breaking scenarios in the case of the next-to-minimal supersymmetric standard model.
In section \ref{sec:models} we discuss the different anomaly mediated supersymmetry breakings in the Minimal Supersymmetric
Standard Model (MSSM) in terms of the parameter spaces for these models and discuss the limits which can be applied due 
to the present data from particle physics experiments and from relic dark matter density assuming the standard 
model of cosmology. In section \ref{sec:models2} similar scenarios are considered in the Next-to-Minimal Supersymmetric
Standard Model (NMSSM) and the corresponding particle and cosmological bounds are discussed. 
In section \ref{sec:cosmologies} we discuss how alternative cosmological models can affect dramatically the bounds on 
the parameter space of the models we have considered while letting unchanged the observable cosmology. Four different 
alternatives to the standard cosmology are discussed which share this behaviour.  
Section \ref{sec:constr} and section \ref{sec:lhc} discuss respectively the constraints implied by these different cosmological 
scenarios and the perspectives at the LHC for a list of benchmark points which are representative of the available parameter space for 
these AMSB models. Our conclusions are given in section \ref{sec:conclusion}.


\section{Anomaly mediated symmetry breaking in the MSSM}
\label{sec:models}
The superconformal Anomaly Mediated Supersymmetry Breaking (AMSB) mechanism \cite{amsb} is one of the most well-known and 
attractive set-ups for supersymmetry breaking.  Supersymmetry breaking effects in the observable 
sector have in this framework a gravitational origin.
Superconformal symmetry is classically preserved in theories without dimensional parameters and it is in general 
broken by the quantum effects.  As anomalies only depend on the low-energy effective theory, the same will be true for 
the soft terms. Usually the AMSB scenario cannot be applied to the MSSM, as it leads to tachyonic sleptons. However the 
presence of an intermediate threshold can displace the soft terms and avoid this problem.
In superconformal gravity one  introduces a chiral superfield playing the role of the compensating multiplet for super-Weyl 
transformations, called the Weyl or conformal compensator. The F-term vacuum expectation value of the conformal 
compensator is turned on by the supersymmetry breaking in the hidden sector and the soft breaking of supersymmetry in the 
visible sector appears through the chiral anomaly supermultiplet. As the soft SUSY breaking terms arise from the anomaly, the 
supersymmetry breaking terms do not dominate at tree-level. Several soft SUSY breaking scenarios 
can be realised starting from this setup. We discuss in the following some of these realisations.

\subsection{Minimal AMSB}

The minimal AMSB (mAMSB) scenario \cite{amsb} has very attractive properties, since the soft SUSY breaking terms are calculated in terms 
of one single parameter, namely the gravitino mass $m_{3/2}$, and the soft terms are renormalisation group invariants which can 
be calculated for any scale choice. However, the AMSB scenarios suffer from the problem that slepton squared masses are found 
to be negative, leading to tachyonic states. A solution to this problem is to consider that the scalar particles acquire a universal mass $m_0$ at the GUT scale, which when added to the AMSB soft SUSY breaking terms, makes them positive. Therefore, the mAMSB model 
relies on only four parameters:
\begin{equation}
 m_0, m_{3/2}, \tan \beta, \mbox{sgn}(\mu) \;.
\end{equation}

This scenario has been thoroughly studied in the literature, but is known to have cosmological consequences incompatible with the 
WMAP observations of the dark matter density \cite{Baer:2010kd}. We perform here an updated analysis of the mAMSB parameter space constraints 
from flavour physics and cosmological relic density. For this study, we generate mass spectra and couplings using Isajet 7.80 
\cite{isajet}. The calculation of flavour observables and the computation of the relic density are performed with SuperIso Relic v3.0 \cite{superiso,superiso_relic}. We use the constraints given in Table~9 of the latest version of the SuperIso manual.

The first flavour observable that we consider here is the branching ratio of $ B \to X_s \gamma$, which has been been thoroughly 
studied in the literature and is still under scrutiny. This observable is very interesting, as its SM contributions only appear at loop 
level, and its theoretical uncertainties as well as the experimental errors are now under control. It provides strong constraints on 
the supersymmetric parameter space, especially for large $\tan \beta$, where it receives large enhancements from its 
supersymmetric contributions. We use the following interval at 95\% C.L. \cite{superiso,Mahmoudi:2007gd}
:
\begin{equation}
 2.16 \times 10^{-4} < \mbox{BR}(B \to X_s \gamma) < 4.93 \times 10^{-4} \;.
\end{equation}

Another interesting observable is the branching fraction of $B_s \to \mu^+ \mu^-$, which is also a loop level observable, and which 
can receive extremely large contributions from SUSY at large $\tan\beta$, and can receive an enhancement of several orders of 
magnitude compared to the SM branching ratio. This decay mode has not yet been observed, and we have at 95\% C.L. \cite{superiso,CDF_bsmumu}
:
\begin{equation}
\mbox{BR}(B_s \to \mu^+ \mu^-) < 4.7 \times 10^{-8} \;.
\end{equation}

\begin{figure}[p!]
\begin{center}
\vspace*{-2.cm}\includegraphics[width=16cm]{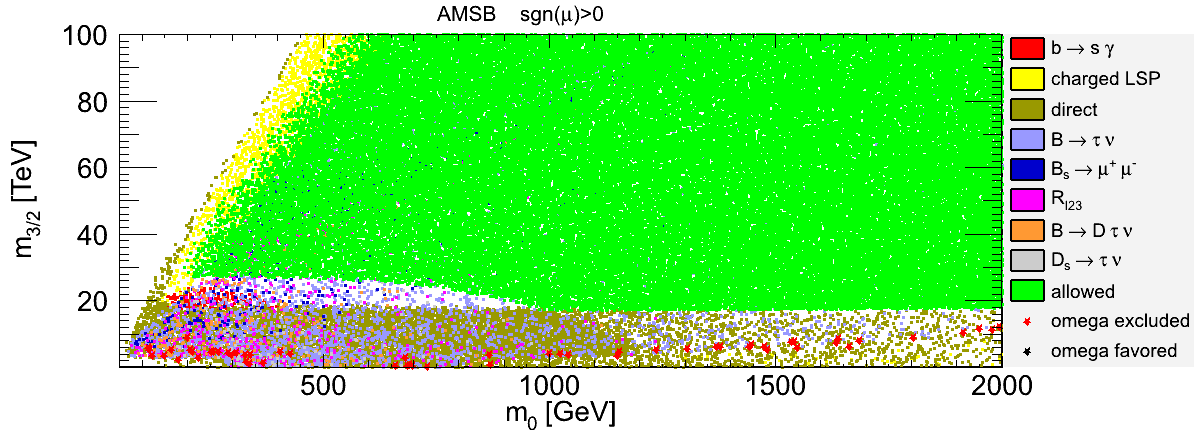}\\
\includegraphics[width=16cm]{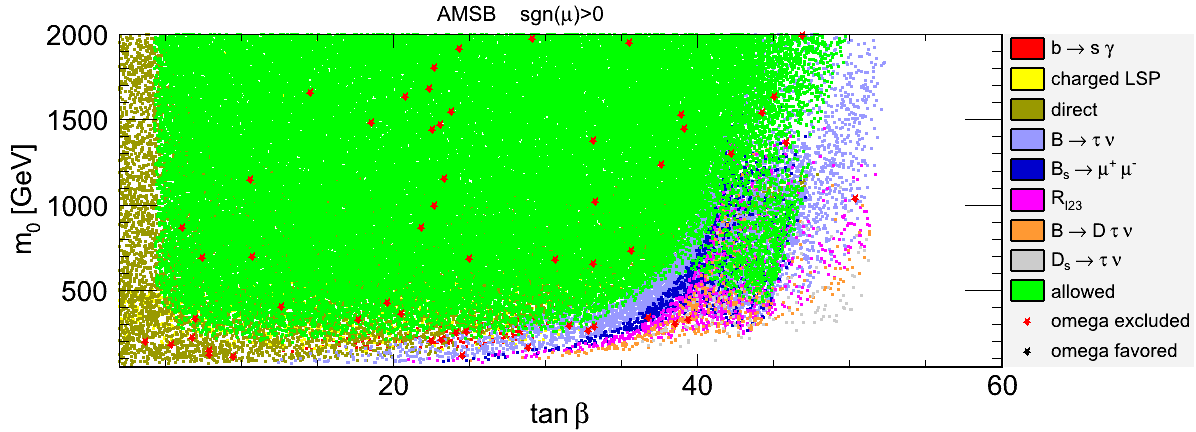}\\
\includegraphics[width=16cm]{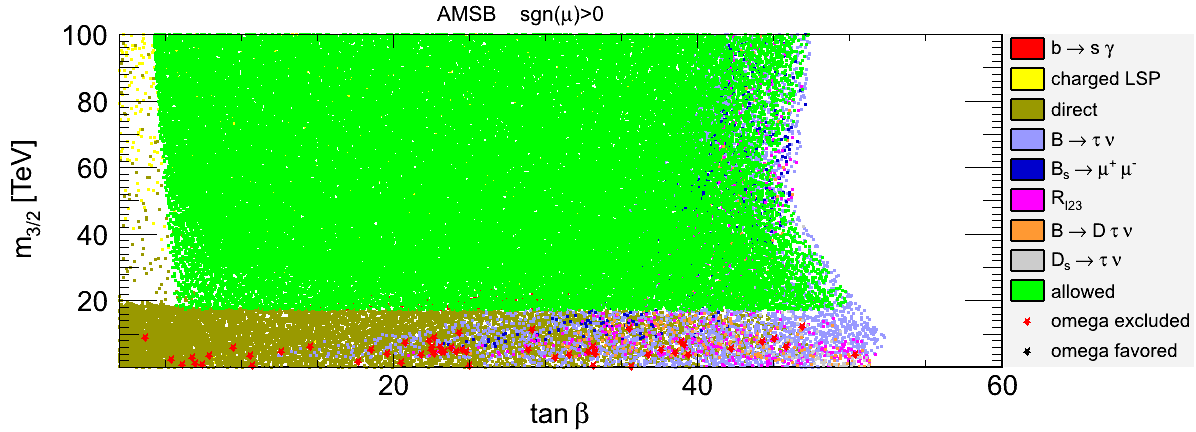}
\end{center}
\caption{\label{fig_amsb} Constraints on the minimal AMSB parameter space. The exclusion regions are plotted in the order given 
in the legend. The red zones are excluded by the inclusive branching ratio of $B \to X_s \gamma$, the yellow ones correspond to 
charged LSP, the olive-green areas are excluded by direct collider constraints, the light blue zones are excluded by 
BR($B\to\tau \nu$), the dark blue zones by BR($B_s\to\mu^+ \mu^-$), the magenta zones by $R_{\ell 23}$, the orange zones by 
BR($B\to D \tau \nu$) and the grey zones by BR($D_s\to \tau \nu$). The green area are in agreement with all the previously 
mentioned constraints. The stars are points favoured by the relic density observable, in red if disfavoured by any other constraints 
and in black if in agreement with all the constraints simultaneously.}
\end{figure}

We also consider a set of tree-level observables which are very sensitive to the charged Higgs mass as well as $\tan\beta$, and 
we use the following 95\% level intervals, which include the theoretical and experimental errors \cite{superiso,Akeroyd:2009tn,Antonelli:2008jg}:
\begin{eqnarray}
0.56 < &\dfrac{\mbox{BR}(B \to \tau \nu)}{\mbox{BR}_{SM}(B \to \tau \nu)}& < 2.70 \;,\\
4.7 \times 10^{-2} < &\mbox{BR}(D_s \to \tau \nu )& < 6.1 \times 10^{-2} \;,\\
0.151 < &\dfrac{\mbox{BR}(B \to D^0 \tau \nu)}{\mbox{BR}(B \to D^0 e \nu)}& < 0.681 \;,\\
0.982 < &\mbox{R}_{\ell23}(K \to \mu \nu) & < 1.018 \;.
\end{eqnarray}
The observable $\mbox{R}_{\ell23}(K \to \mu \nu)$ is related to the decay of $K \to \mu \nu$ and is detailed 
in \cite{Antonelli:2008jg}.
\\
For the relic density constraint, we use the WMAP constraints \cite{Komatsu:2010fb} increased by 10\% of theoretical error to account for the uncertainties in the calculation of the relic density:
\begin{equation}
0.088 < \Omega_{DM} h^2 < 0.123 \;.
\end{equation}
In the following, we disregard the case of negative $\mbox{sgn}(\mu)$ since it is disfavoured by the muon 
anomalous magnetic moment 
constraint, and we scan over the intervals $m_0 \in [0, 2000]$ GeV, $m_{3/2} \in [0, 100]$ TeV and $\tan \beta \in [0, 60]$. Figure 
\ref{fig_amsb} presents projection plots of the parameter space into the possible different planes. The green region in the plots 
corresponds to the parameter zone which is not excluded by flavour constraints or mass limits. The red stars corresponds to points 
leading to a favoured relic density but excluded by other constraints, whereas black stars are favoured by all the presented 
constraints, including the relic density constraint. As can be seen, no black star is visible in these plots, and the whole parameter space presented here is disfavoured either by flavour or direct constraints, or by the relic density constraint which tends to favour the low $m_{3/2}$ region. Disregarding the relic density constraint, a large zone at low $m_{3/2}$ is excluded.

\begin{figure}[p!]
\begin{center}
\includegraphics[width=16cm]{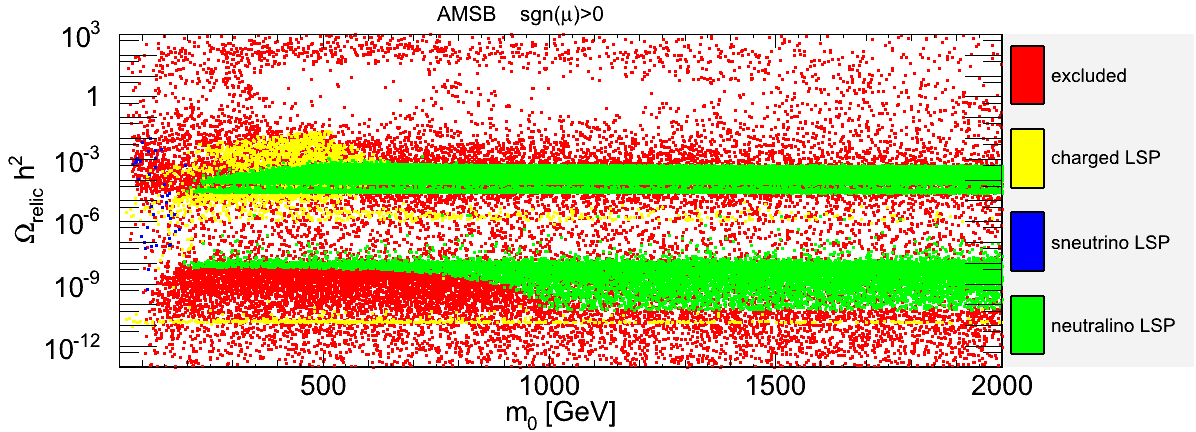}\\
\includegraphics[width=16cm]{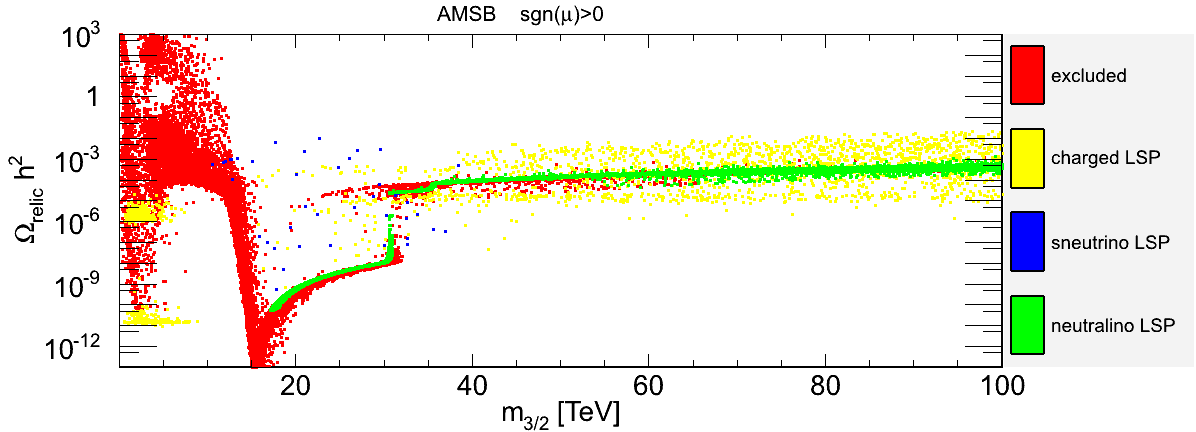}\\
\includegraphics[width=16cm]{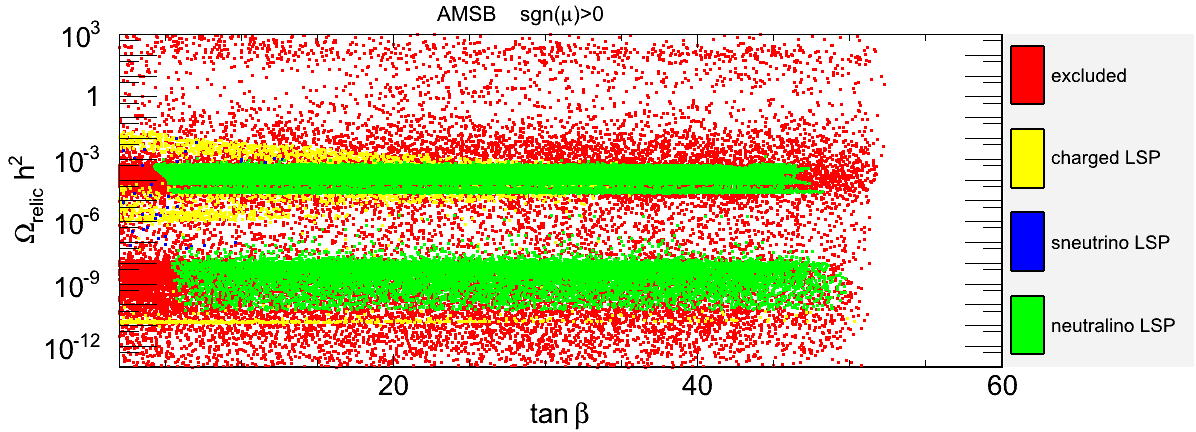}
\end{center}
\caption{\label{fig_amsb_omega}Relic density in function of the AMSB parameters. The green points are favoured by all the 
constraints, the yellow points corresponds to a charged LSP, the blue points correspond to left sneutrino LSP, and the red points 
are excluded by the other constraints (flavour and direct limits).}
\end{figure}

In Fig. \ref{fig_amsb_omega}, we show the relic density values as a function of the AMSB parameters. The green zones 
correspond to regions favoured by the flavour and direct constraints, whereas the other points are either excluded by these 
constraints or by cosmological considerations (charged relic or sneutrino relic, which interact therefore strongly). Two green 
zones clearly appear on the $m_0$ and $\tan\beta$ plots, for $\Omega h^2$ around $10^{-4}$ and $10^{-9}$. These areas 
are far from the WMAP dark matter allowed interval, making the mAMSB scenario disfavoured by the standard cosmology.

\subsection{HCAMSB}

Another possibility to solve the negative slepton squared masses of the original AMSB scenario has been proposed: the hypercharge 
anomaly mediated supersymmetry breaking (HCAMSB) scenario \cite{Dermisek:2007qi}, in which the MSSM resides on a D-brane and the hypercharge 
gaugino mass is generated in a geometrically separated hidden sector \cite{Baer:2009wz}. In this way, additional contribution to 
the gaugino mass $M_1$ is generated, and the large value of $M_1$ then increases the weak scale slepton masses beyond 
tachyonic values, solving the generic AMSB problem \cite{Dermisek:2007qi}.

The HCAMSB scenario has four parameters:
\begin{equation}
 \alpha = \frac{\tilde{M}_1}{m_{3/2}}, m_{3/2}, \tan \beta, \mbox{sgn}(\mu) \;.
\end{equation}
where $\tilde{M}_1$ is the HCAMSB contribution to $M_1$.

In order to study the parameter space, we generate mass spectra and couplings using Isajet 7.80 \cite{isajet} and 
compute the flavour observables and relic density with SuperIso Relic v3.0 \cite{superiso,superiso_relic}. 
We use the constraints described in the previous subsection.

\begin{figure}[p!]
\begin{center}
\includegraphics[width=16cm]{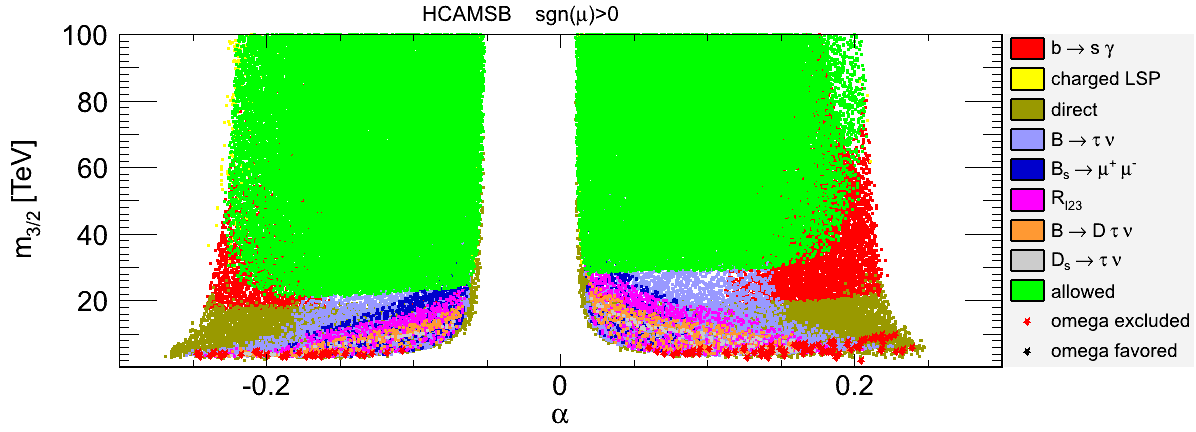}\\
\includegraphics[width=16cm]{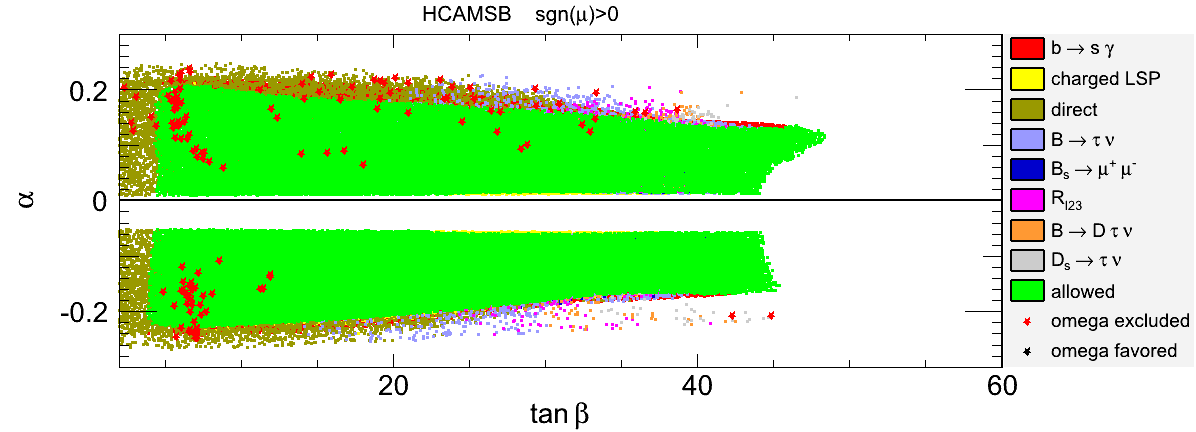}\\
\includegraphics[width=16cm]{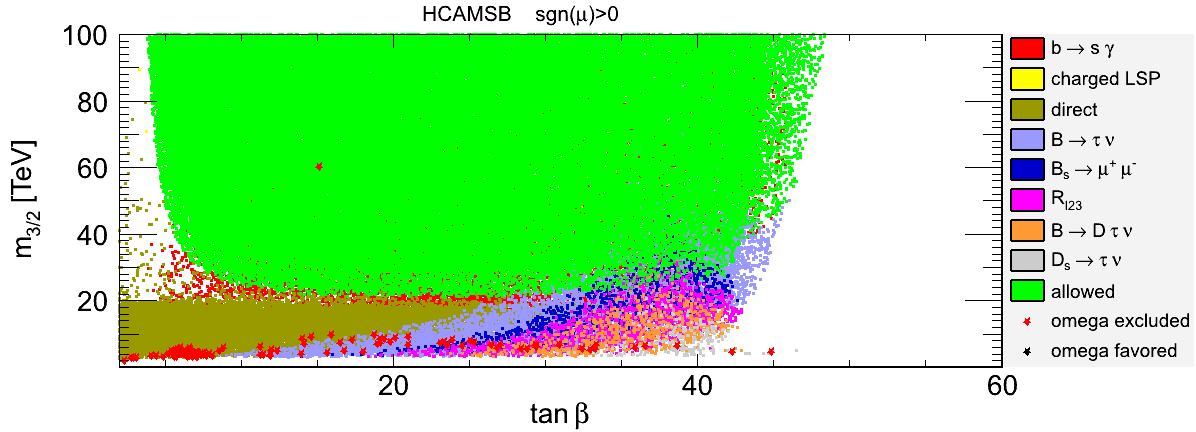}
\end{center}
\caption{\label{fig_hcamsb}Constraints on the HCAMSB parameter space. The colour codes are the same as 
in Fig. \ref{fig_amsb}.}
\end{figure}

In Fig. \ref{fig_hcamsb}, we scan over the whole parameter space, and project the results in two-dimensional planes. The results are somehow similar to those of the mAMSB scenario: the constraints exclude low $m_{3/2}$ values. A large part of the parameter space is favoured by flavour and direct constraints, but unfortunately no part of the parameter space respects at the same time the relic density and the other constraints, so that HCAMSB is also disfavoured by the standard cosmology.

\begin{figure}[p!]
\begin{center}
\includegraphics[width=16cm]{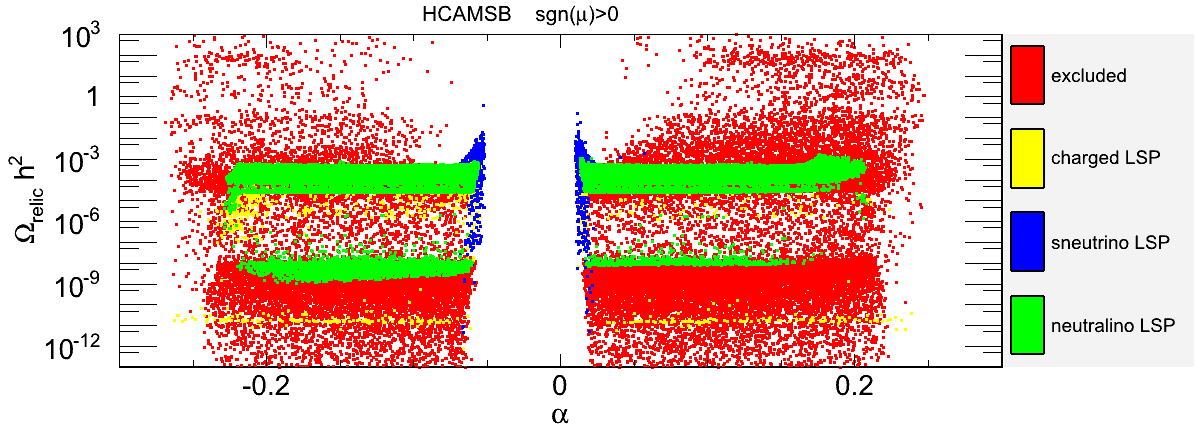}\\
\includegraphics[width=16cm]{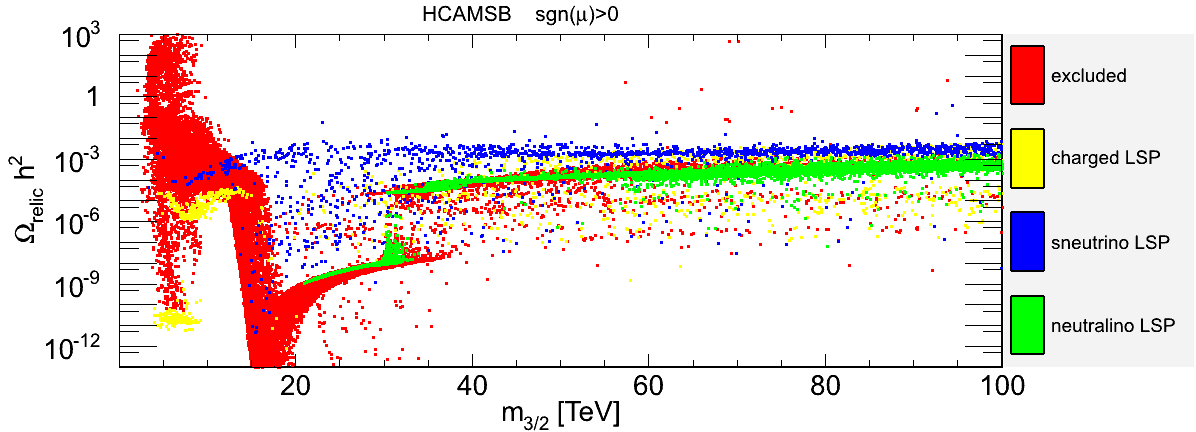}\\
\includegraphics[width=16cm]{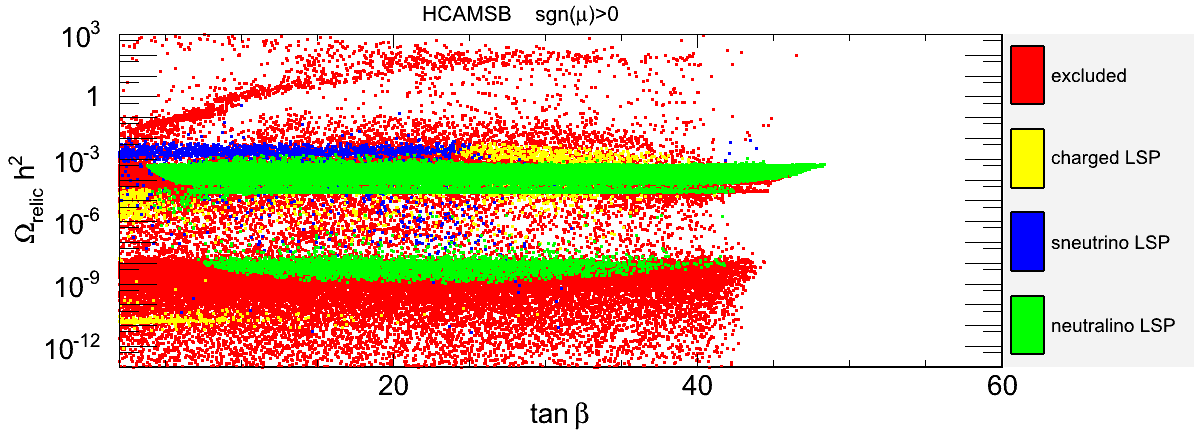}
\end{center}
\caption{\label{fig_hcamsb_omega}Relic density in function of the HCAMSB parameters. The colour codes are the same as 
in Fig. \ref{fig_amsb_omega}.}
\end{figure}

In Fig. \ref{fig_hcamsb_omega}, we show the relic density values as a function of the HCAMSB parameters. 
Again, the results are similar to those of the mAMSB scenario, with two distinct zones which are not excluded by 
flavour and direct constraints, corresponding to a relic density $\Omega h^2$ around $10^{-4}$ and $10^{-8}$.

\subsection{MMAMSB}

Contrary to the two previous AMSB scenarios, the Mixed Modulus Anomaly mediated SUSY breaking (MMAMSB) scenario \cite{Choi:2005uz} provides viable dark matter candidates, in addition to solving the negative slepton mass problem naturally. This scenario is based on type-IIB 
superstrings with stabilised moduli \cite{Baer:2006id}. In this scenario, an interesting result is that the soft SUSY breaking terms 
receive comparable contributions from both anomaly and modulus, resulting in positive slepton masses. We examine 
here the minimal MMAMSB scenario\footnote{{We note that in the general MMAMSB, there are two more 
parameters, $n_{i}$ and $l_{\alpha}$, which represent respectively the modular weights of visible sector of the matter fields and 
the gauge kinetic function, and which can modify the mass spectra.}}, which relies on four parameters:
\begin{equation}
 \alpha , m_{3/2}, \tan \beta, \mbox{sgn}(\mu) \;.
\end{equation}
$\alpha$ here parametrises the relative contributions of modulus mediation and anomaly mediation to the soft breaking terms: the 
largest $\alpha$ is, the more mediation comes from modulus \cite{Choi:2005uz}.

\begin{figure}[p!]
\begin{center}
\includegraphics[width=16cm]{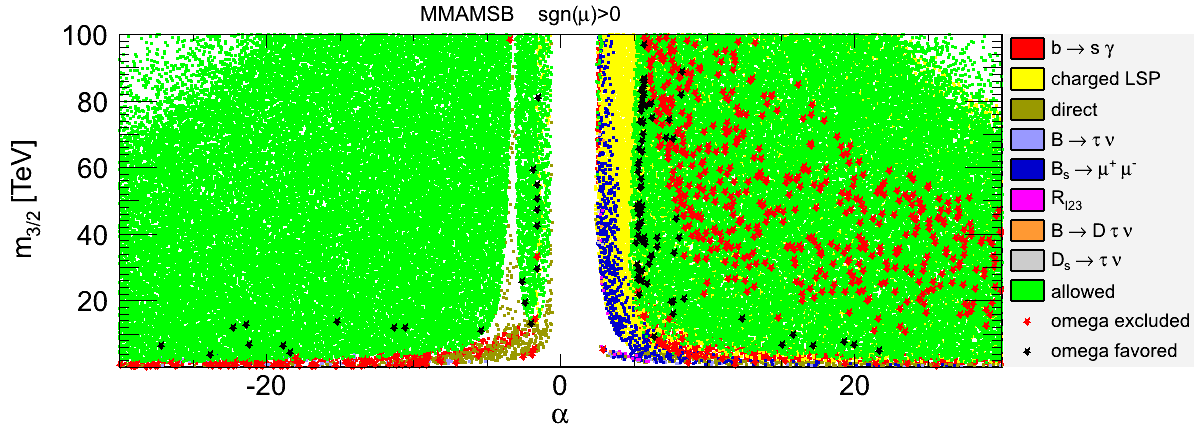}\\
\includegraphics[width=16cm]{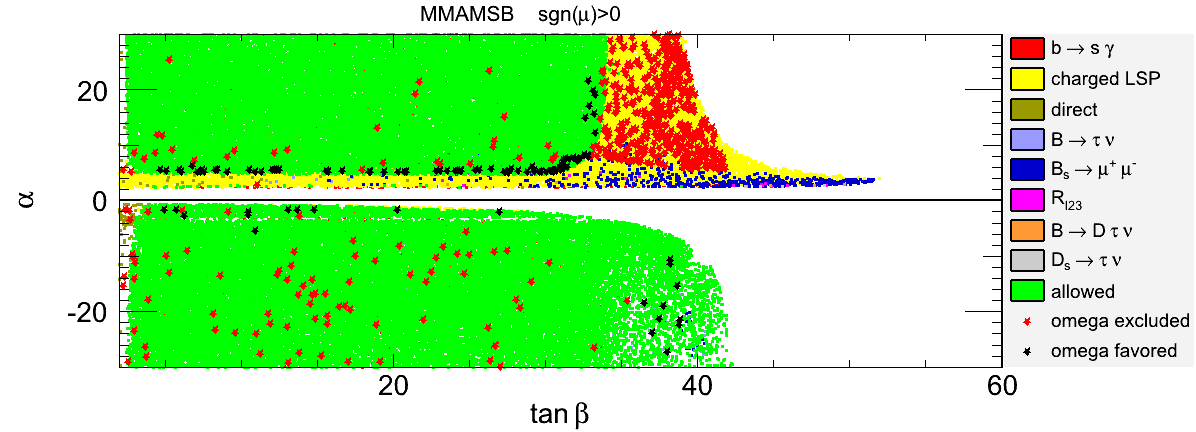}\\
\includegraphics[width=16cm]{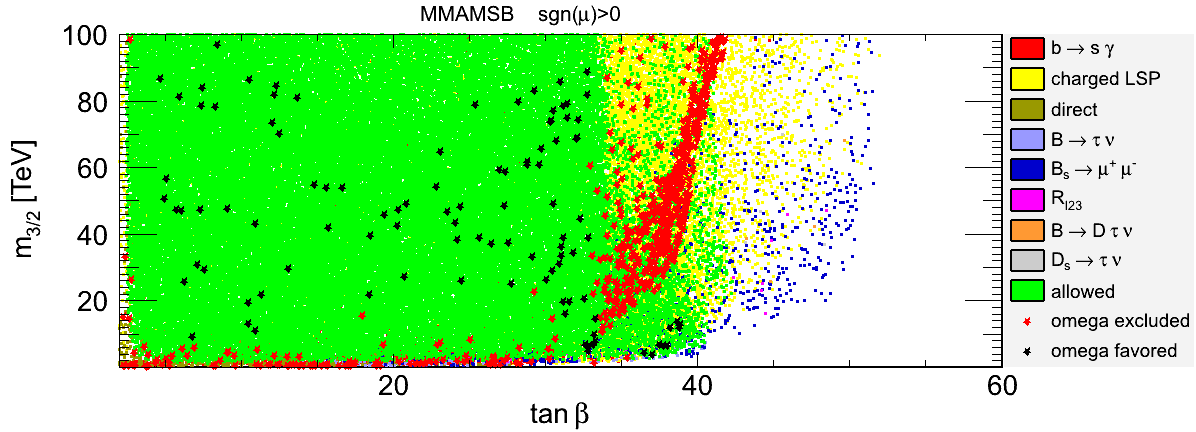}
\end{center}
\caption{\label{fig_mmamsb}Constraints on the MMAMSB parameter space. The colour codes are the same as 
in Fig. \ref{fig_amsb}.}
\end{figure}

In Fig. \ref{fig_mmamsb}, the parameter space is scanned over, and it is projected in two-dimensional planes. 
The resulting plots are different from those of mAMSB and HCAMSB. Indeed, a large part of parameter space escapes 
the flavour and direct constraints, and zones around $\alpha \sim 0$ or at low $m_{3/2}$ also fulfil the relic density constraint.

\begin{figure}[p!]
\begin{center}
\includegraphics[width=16cm]{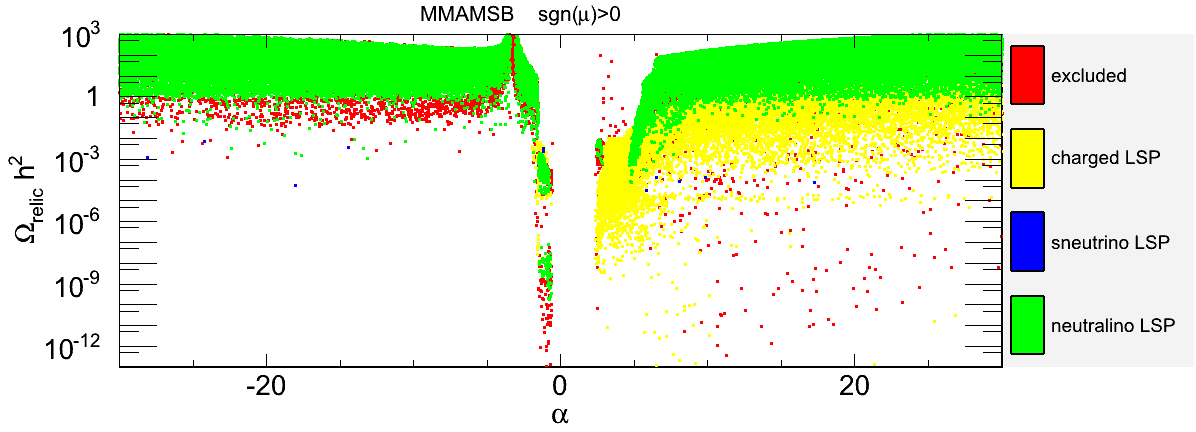}\\
\includegraphics[width=16cm]{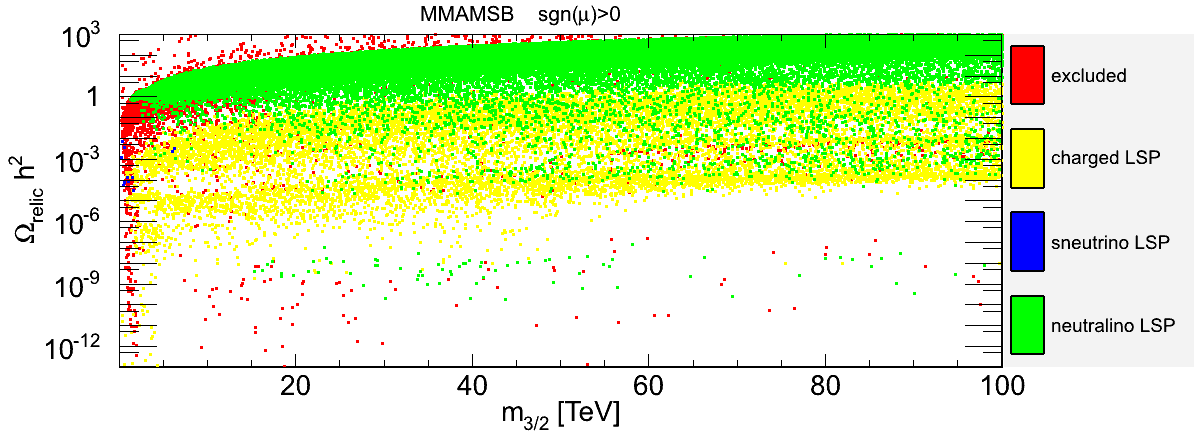}\\
\includegraphics[width=16cm]{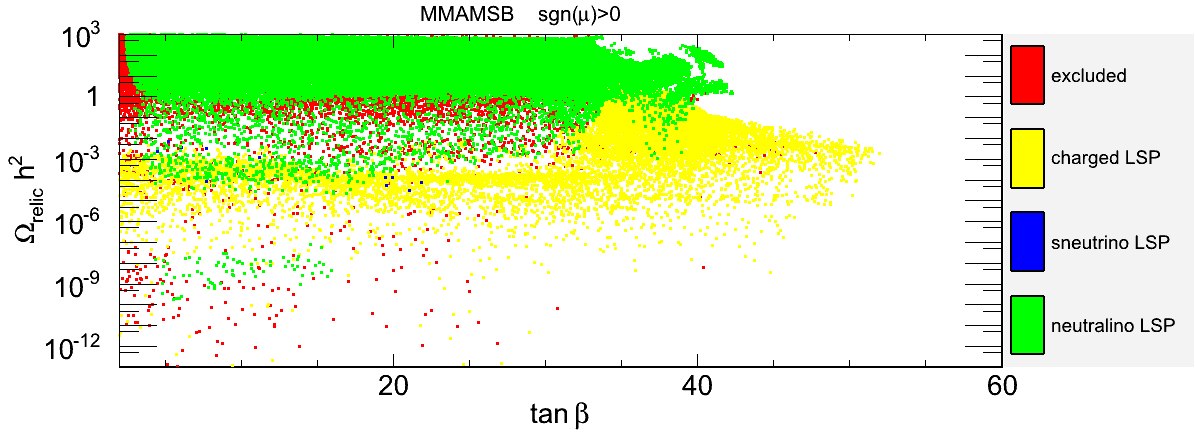}
\end{center}
\caption{\label{fig_mmamsb_omega}Relic density in function of the MMAMSB parameters. The colour codes are the same 
as in Fig. \ref{fig_amsb_omega}.}
\end{figure}

In Fig. \ref{fig_mmamsb_omega}, we present the relic density values in function of the MMAMSB parameters. The relic density of points
favoured by flavour and direct constraints takes values between $10^{-9}$ and  $10^3$. We can notice however that most of the 
points are in the interval $[1,10^3]$, and some points fit in the WMAP interval. For this reason, MMAMSB is a scenario which can 
appear as attractive, as it fulfils simultaneously the standard cosmology and particle physics constraints.

\section{Anomaly mediated symmetry breaking in the NMSSM}
\label{sec:models2}
An interesting extension of the MSSM is the NMSSM, which brings a solution to the $\mu$-problem \cite{Maniatis:2009re}. It has an extended Higgs 
sector involving additional Higgs bosons, modifying the relic density and flavour physics constraints. 
Moreover, the couplings being modified, the NMSSM can escape direct constraints, and new parameter zones can be allowed by 
the constraints used in the previous section.

We consider here the simplest version of the NMSSM, where the term $\mu \hat H_u \cdot \hat H_d$ of the MSSM superpotential 
is replaced by
\begin{equation}
 \lambda \hat H_u \cdot \hat H_d \hat S + \frac\kappa 3 \hat S^3 \;,
\end{equation}
in order for the superpotential to be scale invariant. The soft breaking terms
\begin{equation}
 m_{H_u}^2 |H_u|^2 + m_{H_d}^2 |H_d|^2 + m_S^2 |S|^2 + \bigl(\lambda A_\lambda S H_u\cdot H_d 
 + \frac13 \kappa A_\kappa S^3 + \mbox{h.c.} \bigr) \;,
\end{equation}
are {\it a priori} independent. Using the minimisation conditions for the potential, the scalar mass parameters $m_{H_{u,d}}$ 
can be replaced by the vacuum expectation values of the doublet $v_u$ and $v_d$, with
\begin{equation}
 v_u^2 + v_d^2 = v^2 \approx (174 \mbox{ GeV})^2 \;\;, \qquad \tan\beta = \frac{v_u}{v_d} \;. 
\end{equation}
The singlet field mass parameter can also be replaced by the singlet expectation value $v_s$. Expanding the singlet field $S$ 
around $v_s$ gives rise to an effective parameter $\mu_{eff}= \lambda v_s$.

One can also define an effective doublet mass such as
\begin{equation}
 m_A^2 \equiv \frac{\lambda v_s}{\sin\beta \cos\beta} (A_\lambda + \kappa v_s) \;.
\end{equation}

Once the MSSM-like parameters (and in particular $\mu_{eff}$) have been fixed by the specification of the AMSB scenario, we 
are left with four additional independent NMSSM-specific parameters:
\begin{equation}
 \lambda, \kappa, A_\kappa, M_A \;.
\end{equation}
We scan over the intervals $\lambda \in [-0.7, 0.7]$ GeV, $\kappa \in [-0.7, 0.7]$, $A_\kappa \in [-2000, 2000]$ GeV and 
$M_A \in [5,1000]$. We review in the following the differences between the different AMSB scenarios when applied to the 
NMSSM.

\subsection{mNAMSB}

Minimal NMSSM-AMSB (mNAMSB) parameter points are generated here using NMSSMTools 2.3.4 \cite{NMSSMTools}, 
and flavour constraints and relic density are computed with SuperIso Relic v3.0 \cite{superiso,superiso_relic}.

\begin{figure}[p!]
\begin{center}
\includegraphics[width=16cm]{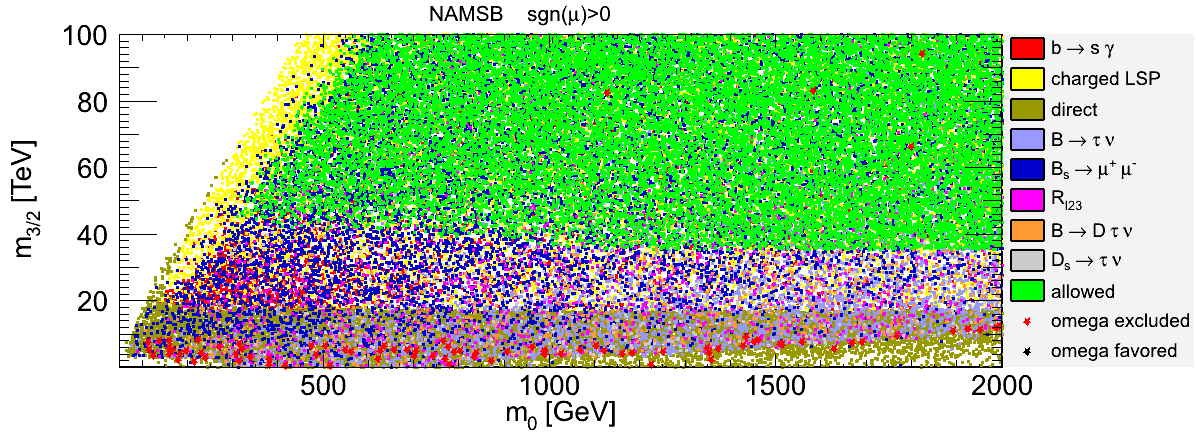}\\
\includegraphics[width=16cm]{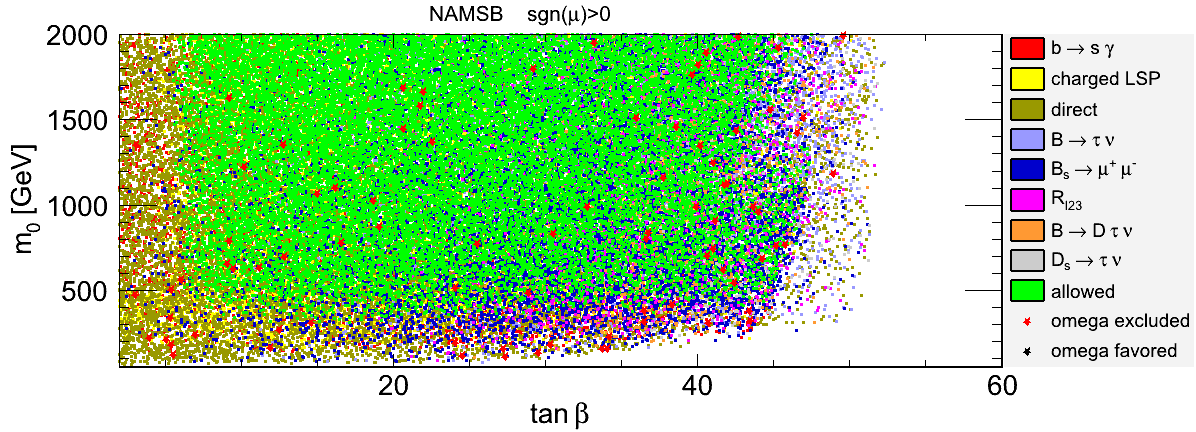}\\
\includegraphics[width=16cm]{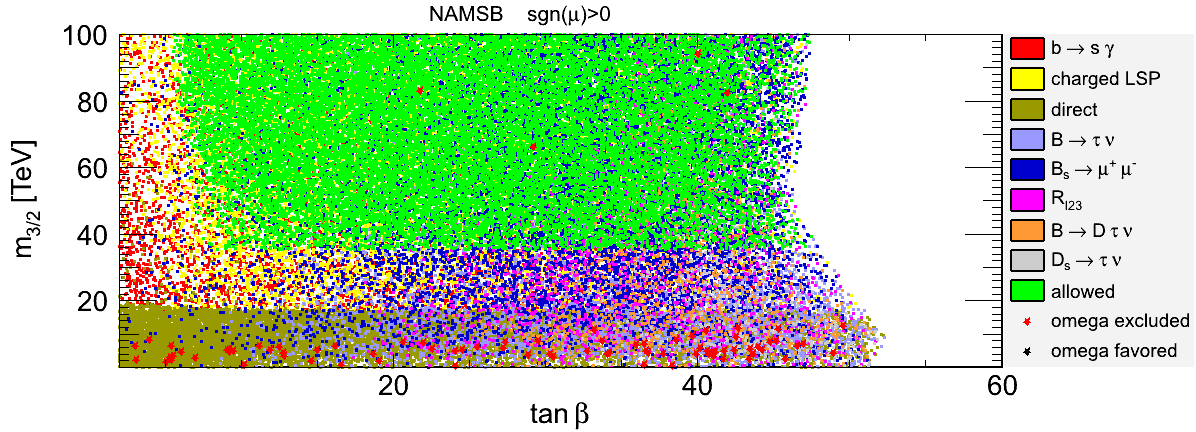}
\end{center}
\caption{\label{fig_namsb}Constraints on the mNAMSB parameter space. The colour codes are the same as in 
Fig. \ref{fig_amsb}.}
\end{figure}

In Fig. \ref{fig_namsb}, the parameter space of mNAMSB is presented. As in Fig. \ref{fig_amsb}, no point satisfy simultaneous 
flavour, direct and relic density constraints. The direct limits are less constraining than in the MSSM, but the flavour constraints 
are stronger and exclude a larger part of the NAMSB parameter space in comparison with AMSB.

\begin{figure}[p!]
\begin{center}
\includegraphics[width=16cm]{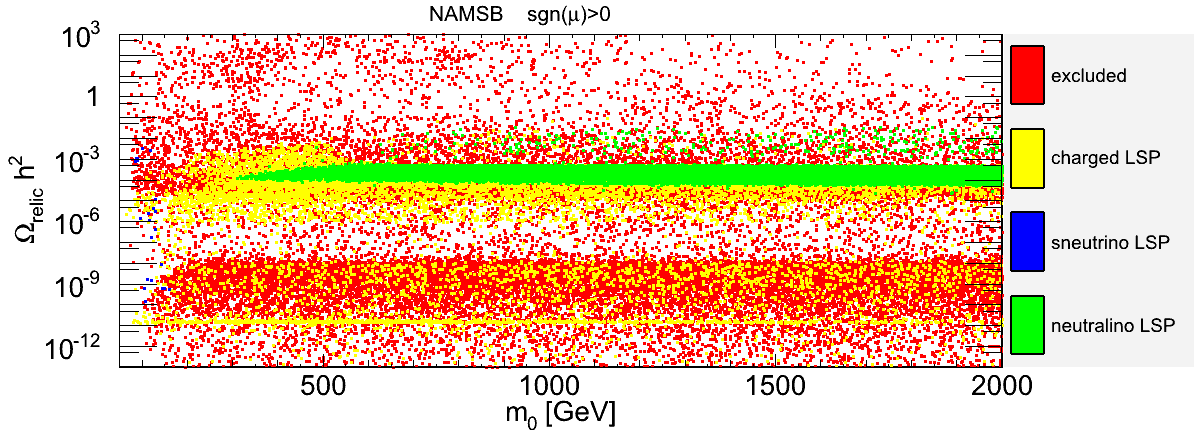}\\
\includegraphics[width=16cm]{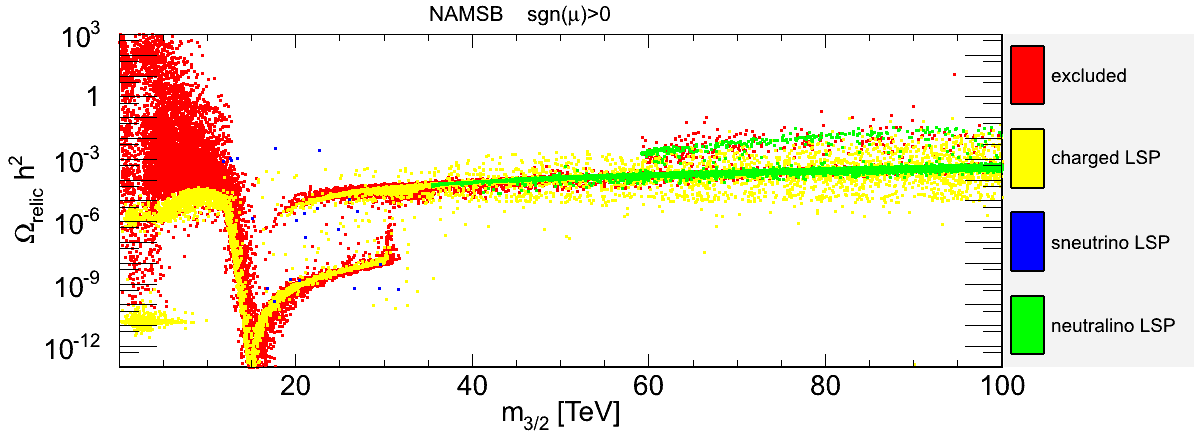}\\
\includegraphics[width=16cm]{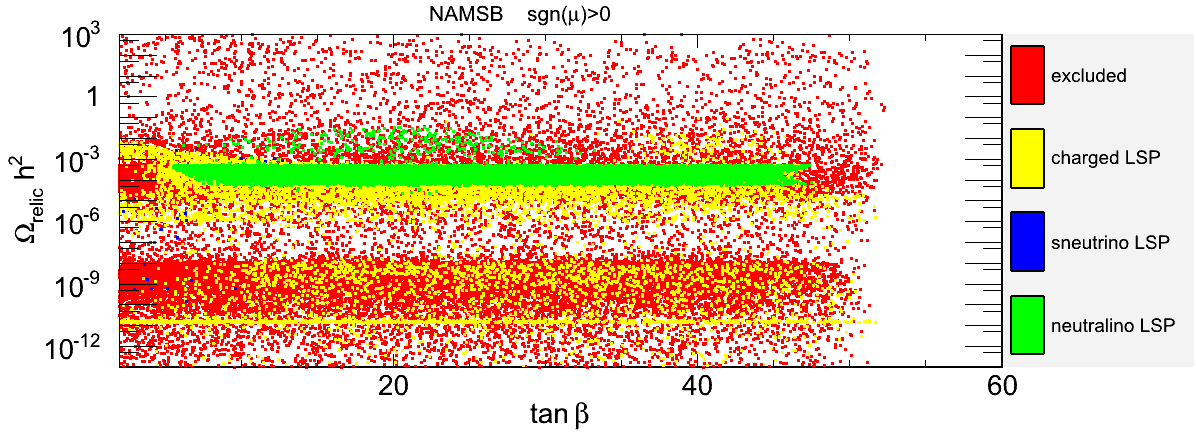}
\end{center}
\caption{\label{fig_namsb_omega}Relic density in function of the mNAMSB parameters. The colour codes are the same as 
in Fig. \ref{fig_amsb_omega}.}
\end{figure}

Fig. \ref{fig_namsb_omega} reveals more differences between the mAMSB and mNAMSB models. First, the zone not 
excluded by flavour and direct constraints having a relic density around $10^{-9}$ does not exist in the NAMSB, a new 
green zone appears for $m_{3/2} > 60$, and its relic density around $10^{-2}$ is much closer to the WMAP constraint. 
However, as in the mAMSB scenario, the mNAMSB model is globally disfavoured by the standard cosmology.

\subsection{NHCAMSB}

\begin{figure}[p!]
\begin{center}
\includegraphics[width=16cm]{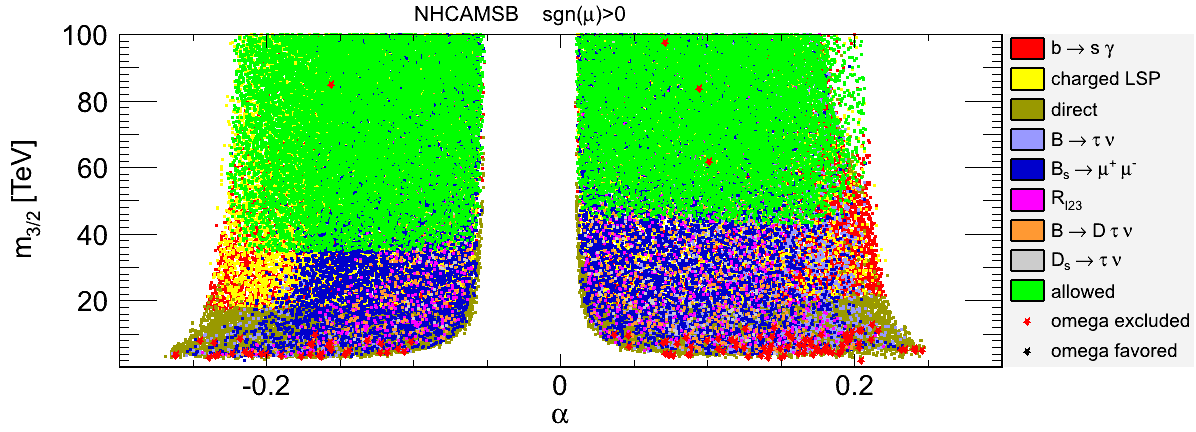}\\
\includegraphics[width=16cm]{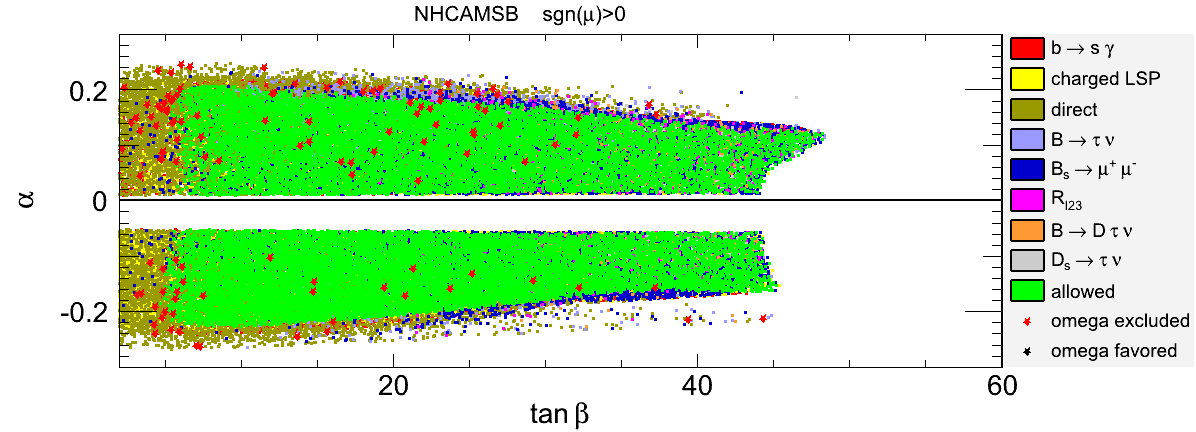}\\
\includegraphics[width=16cm]{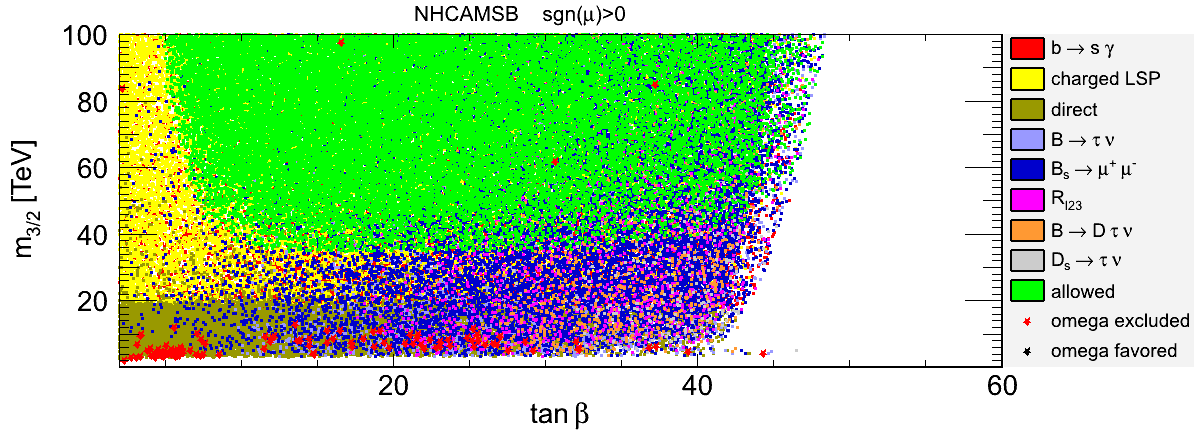}
\end{center}
\caption{\label{fig_nhcamsb}Constraints on the NHCAMSB parameter space. The colour codes are the same as in 
Fig. \ref{fig_amsb}.}
\end{figure}

We generate the NMSSM-HCAMSB (NHCAMSB) parameter points  using NMSSMTools 2.3.4 \cite{NMSSMTools}. 
The obtained constraints are shown in Fig. \ref{fig_nhcamsb}. Again, we see that no parameter point satisfies at the 
same time the relic density constraint and the direct and flavour constraints. Similarly to the mNAMSB scenario, 
NHCAMSB is less constrained by the direct mass limits, but is more excluded by the flavour constraints.

\begin{figure}[p!]
\begin{center}
\includegraphics[width=16cm]{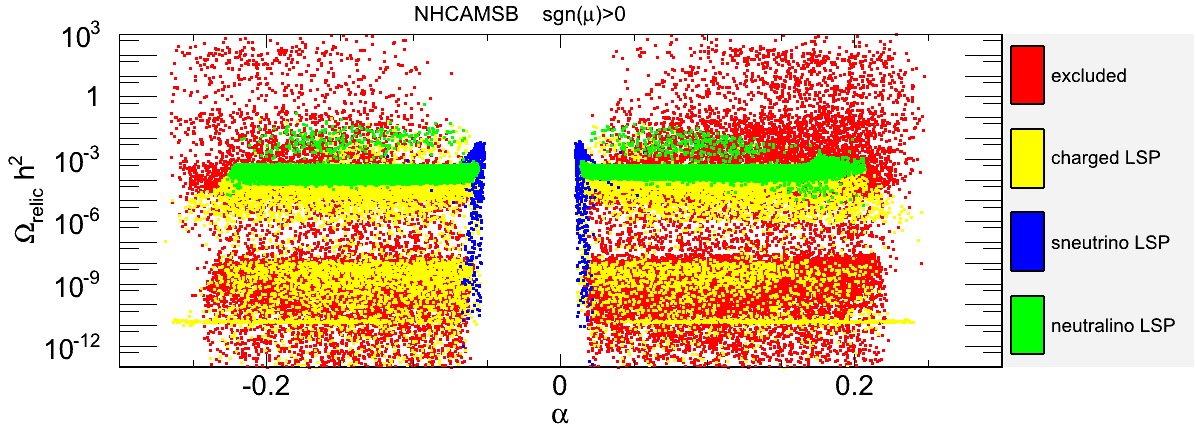}\\
\includegraphics[width=16cm]{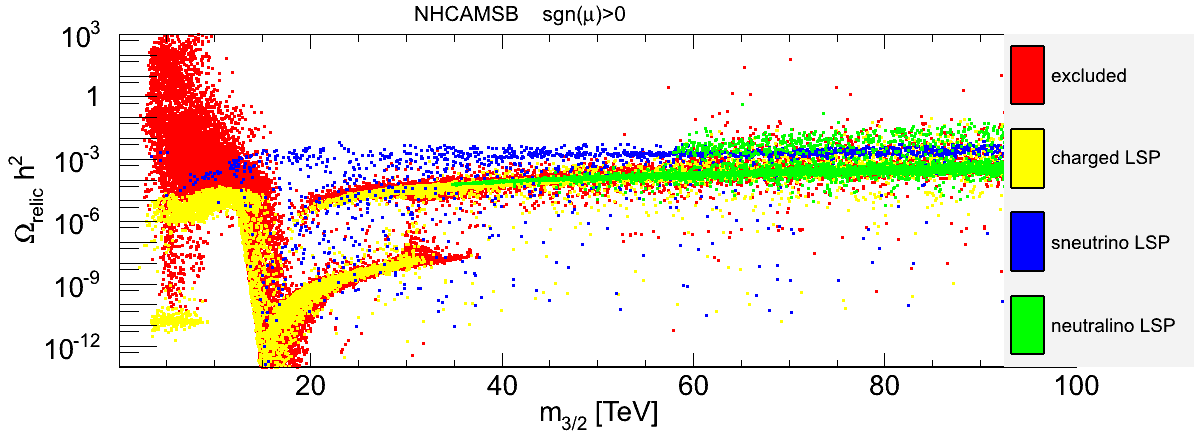}\\
\includegraphics[width=16cm]{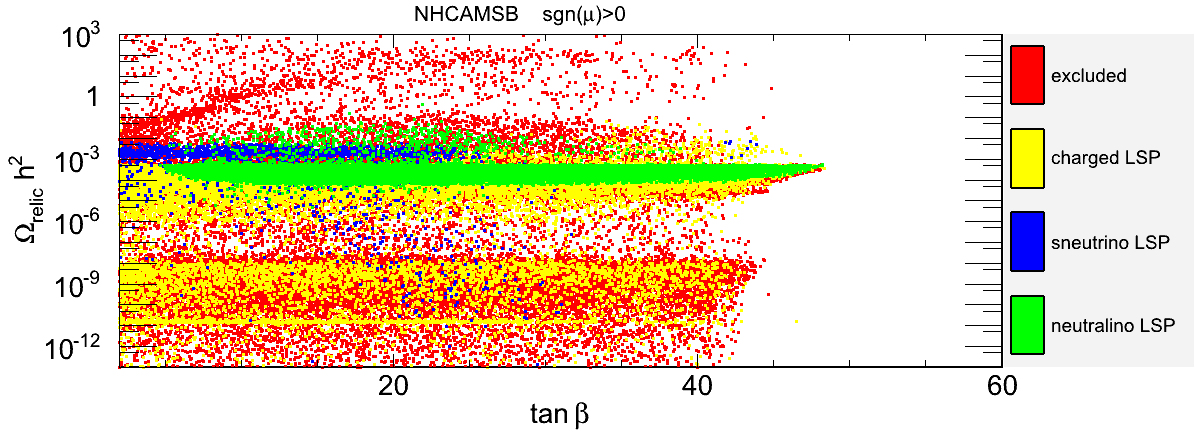}
\end{center}
\caption{\label{fig_nhcamsb_omega}Relic density in function of the NHCAMSB parameters. The colour codes are the 
same as in Fig. \ref{fig_amsb_omega}.}
\end{figure}

Fig. \ref{fig_nhcamsb_omega} shows the relic density in function of the different parameters. The green region that exists 
in the HCAMSB scenario for a relic density around $10^{-8}$ disappears in the NHCAMSB, and a new region opens up 
around $10^{-2}$. However, as the mNAMSB scenario, the NHCAMSB scenario remains also disfavoured by the standard cosmology.

\subsection{NMMAMSB}

\begin{figure}[p!]
\begin{center}
\includegraphics[width=16cm]{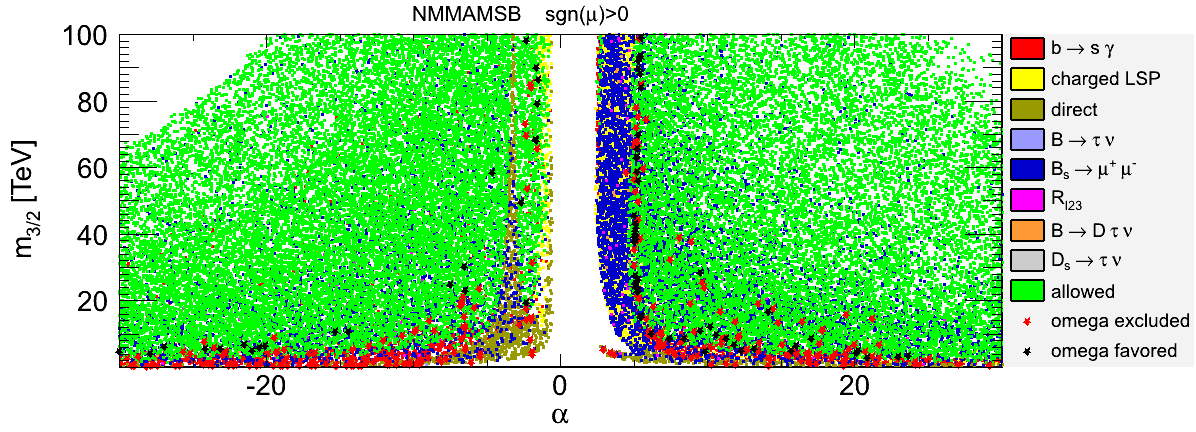}\\
\includegraphics[width=16cm]{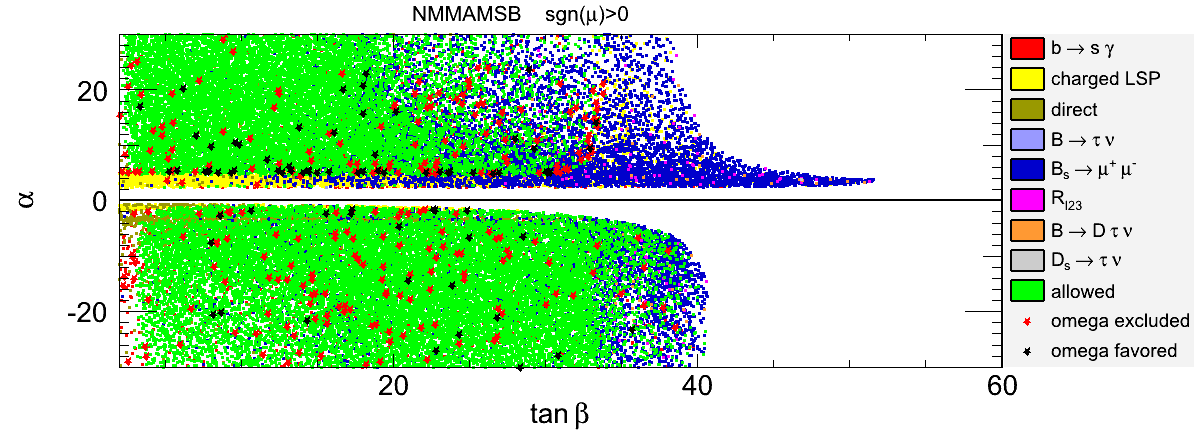}\\
\includegraphics[width=16cm]{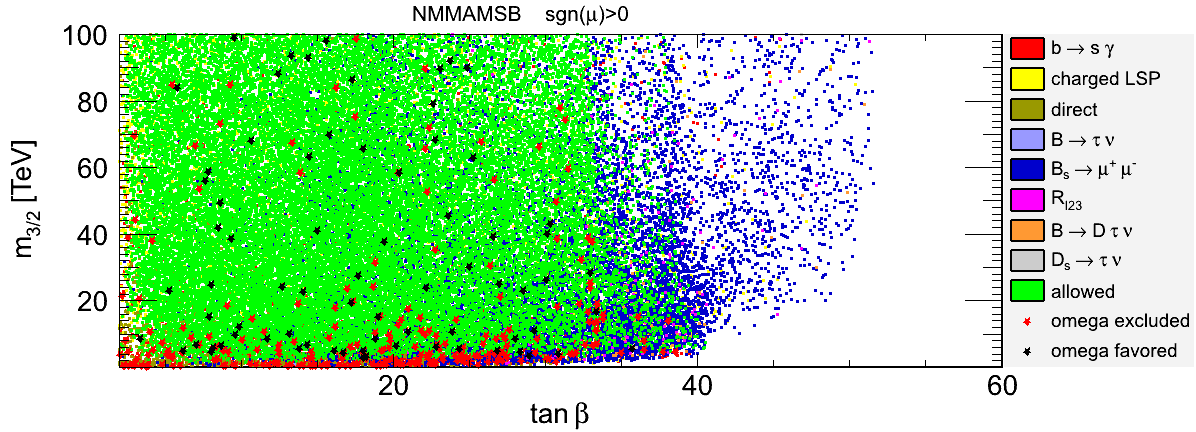}
\end{center}
\caption{\label{fig_nmmamsb}Constraints on the NMMAMSB parameter space. The colour codes are the same as in 
Fig. \ref{fig_amsb}.}
\end{figure}

The NMSSM-MMAMSB (NMMAMSB) scenario leads to similar results as the MMAMSB scenario. As can be seen in 
Fig. \ref{fig_nmmamsb}, there exists many points satisfying all the constraints, including relic density, especially for 
values of $\alpha$ near to 0.

\begin{figure}[p!]
\begin{center}
\includegraphics[width=16cm]{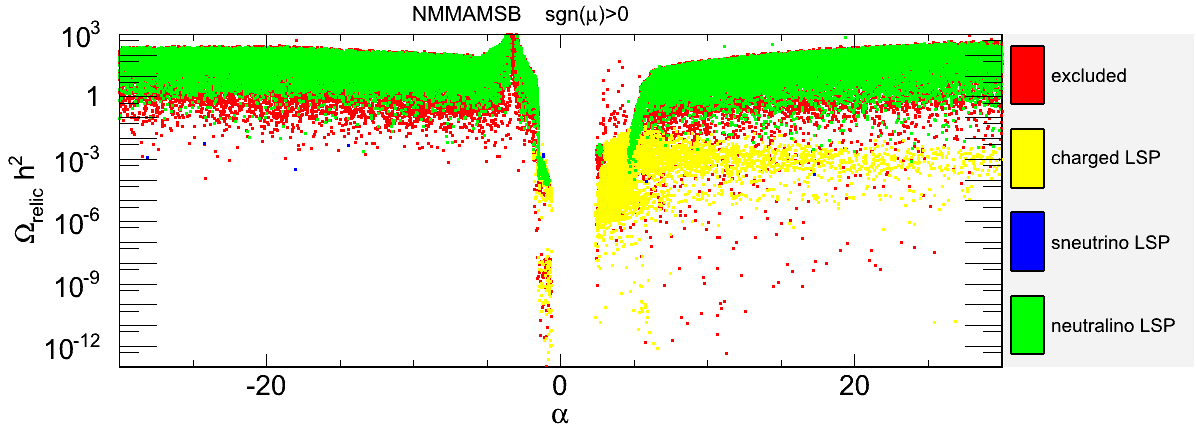}\\
\includegraphics[width=16cm]{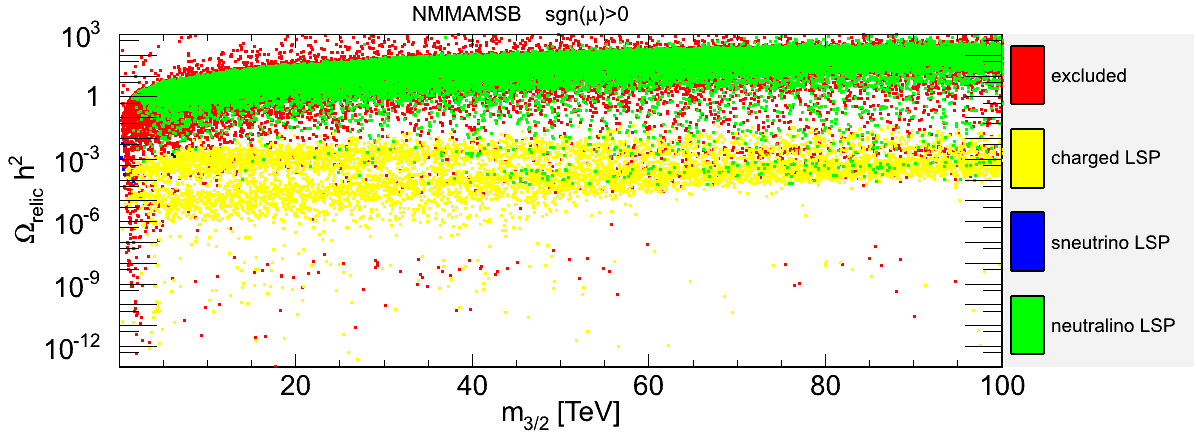}\\
\includegraphics[width=16cm]{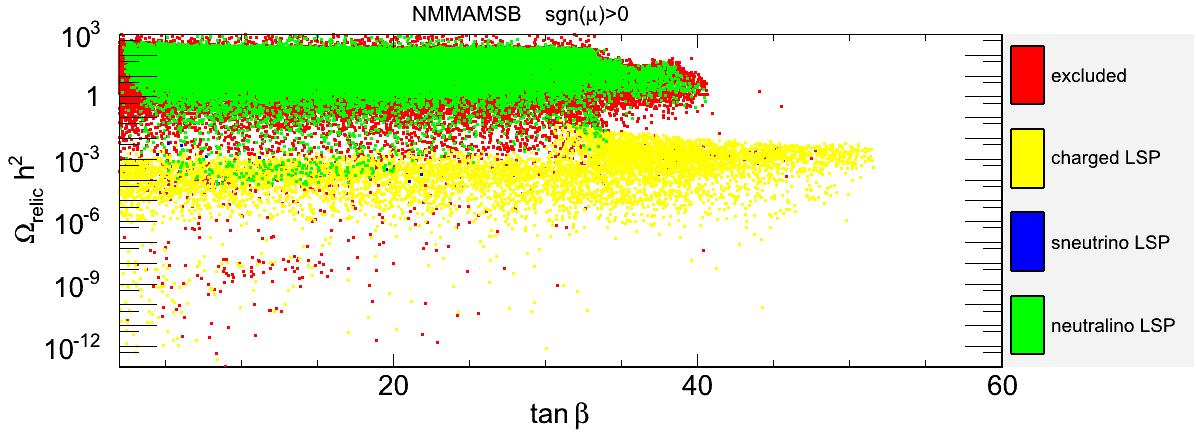}
\end{center}
\caption{\label{fig_nmmamsb_omega}Relic density as a function of the NMMAMSB parameters. The colour codes are the 
same as in Fig. \ref{fig_amsb_omega}.}
\end{figure}

Fig. \ref{fig_nmmamsb_omega} reveals a difference, as the calculated relic density takes values between $10^{-4}$ and 
$10^{3}$, which is more restrictive in comparison to the MSSM case.

\section{AMSB and relic density in alternative cosmology}
\label{sec:cosmologies}
\begin{figure}[p!]
\begin{center}
\hspace*{-4.8cm}Point A \hspace*{6.3cm} Point B\\
\includegraphics[width=8cm]{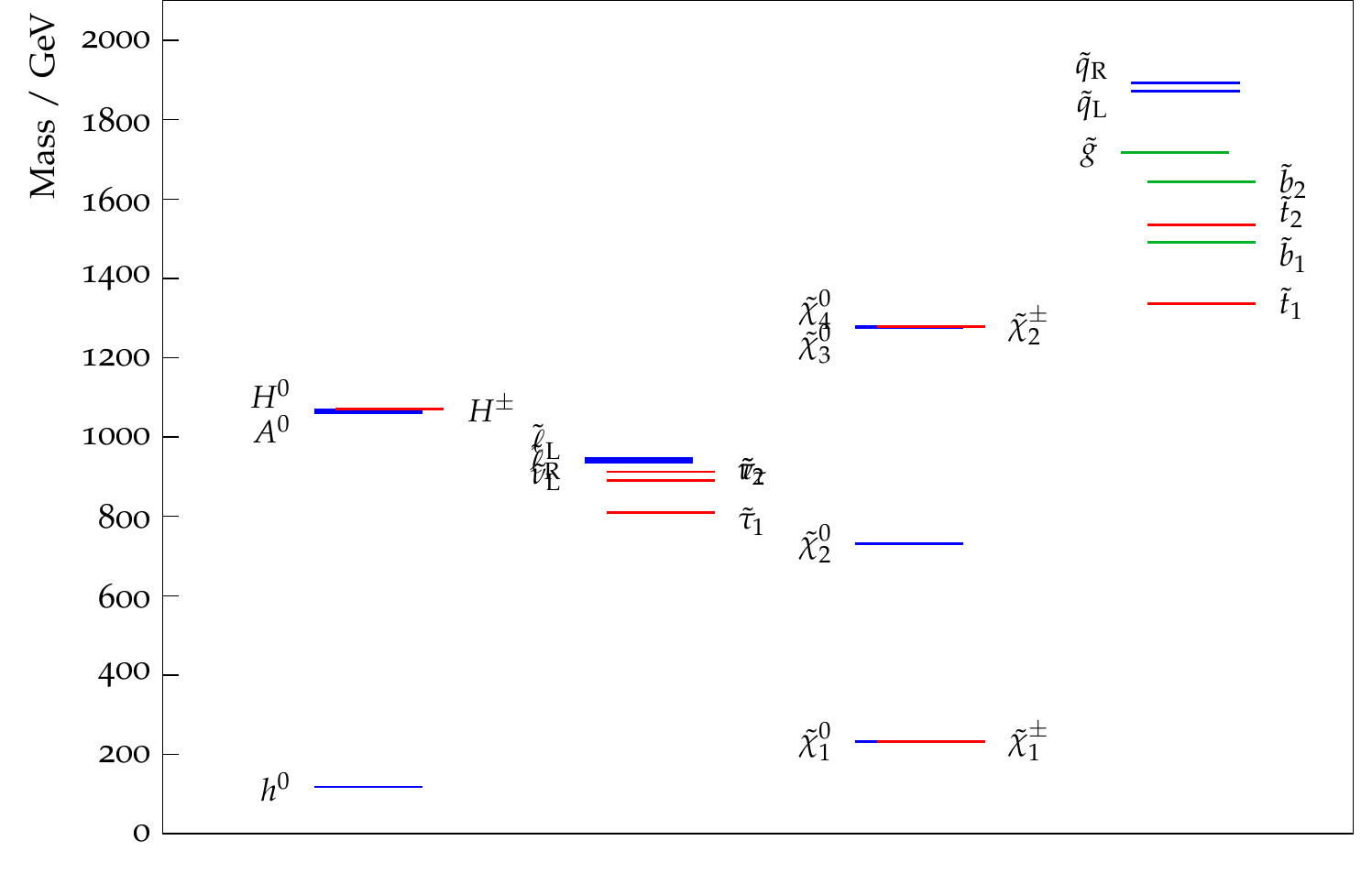}\includegraphics[width=8cm]{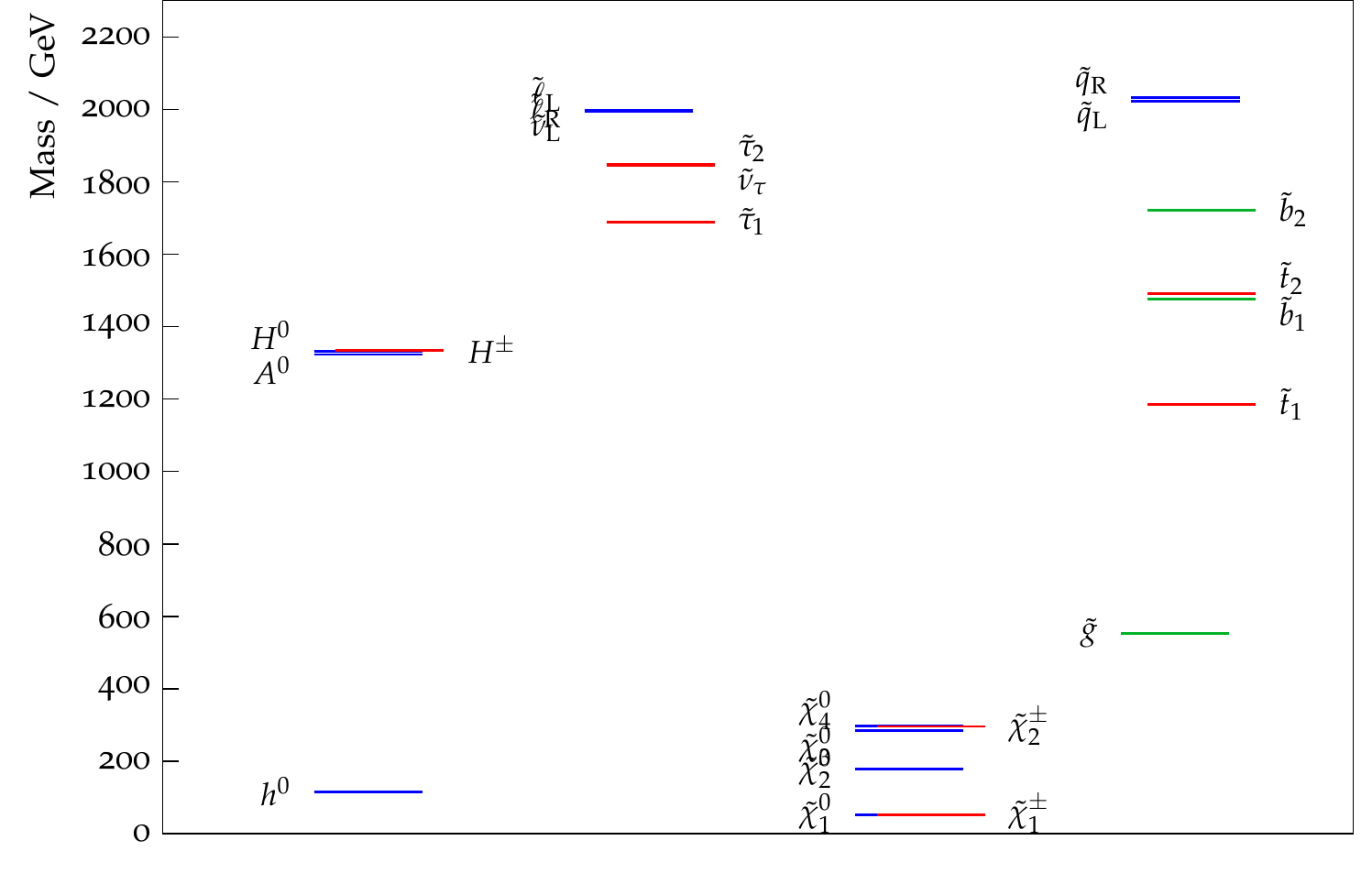}\\
\hspace*{-4.7cm}Point C \hspace*{6.5cm} Point D\\
\includegraphics[width=8cm]{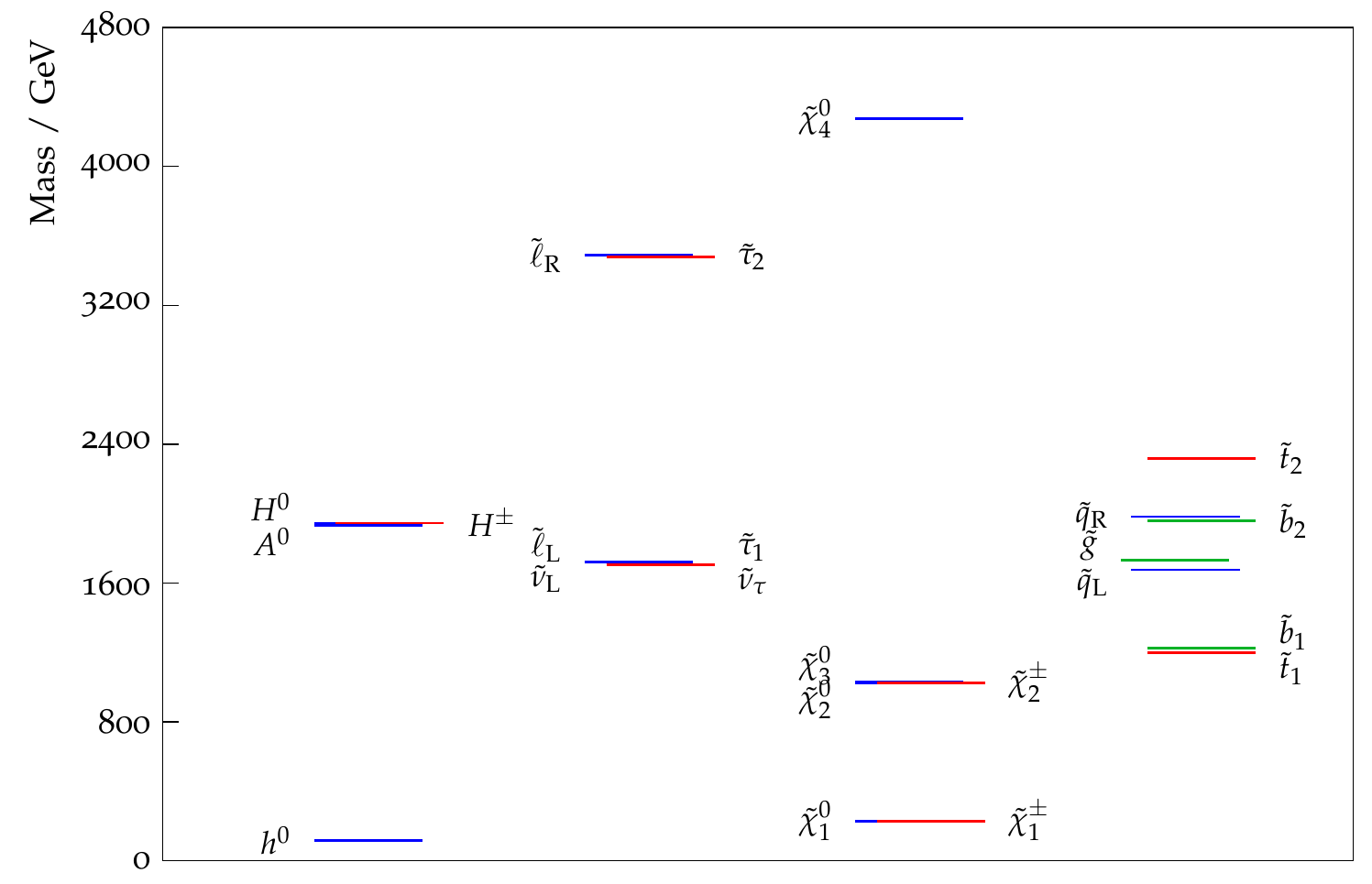} \includegraphics[width=8cm]{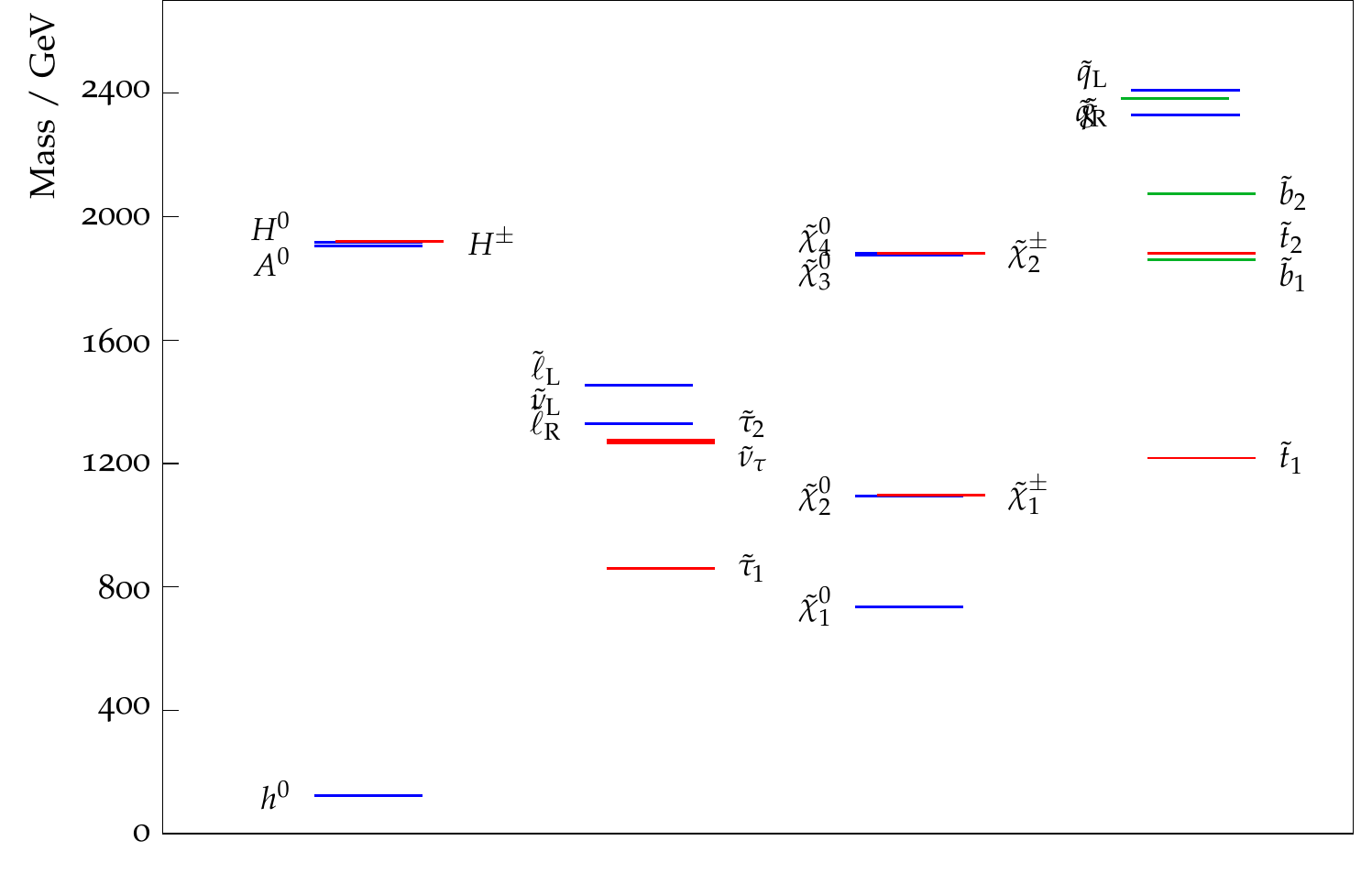}\\
\hspace*{-4.8cm}Point E \hspace*{6.5cm} Point F\\
\includegraphics[width=8cm]{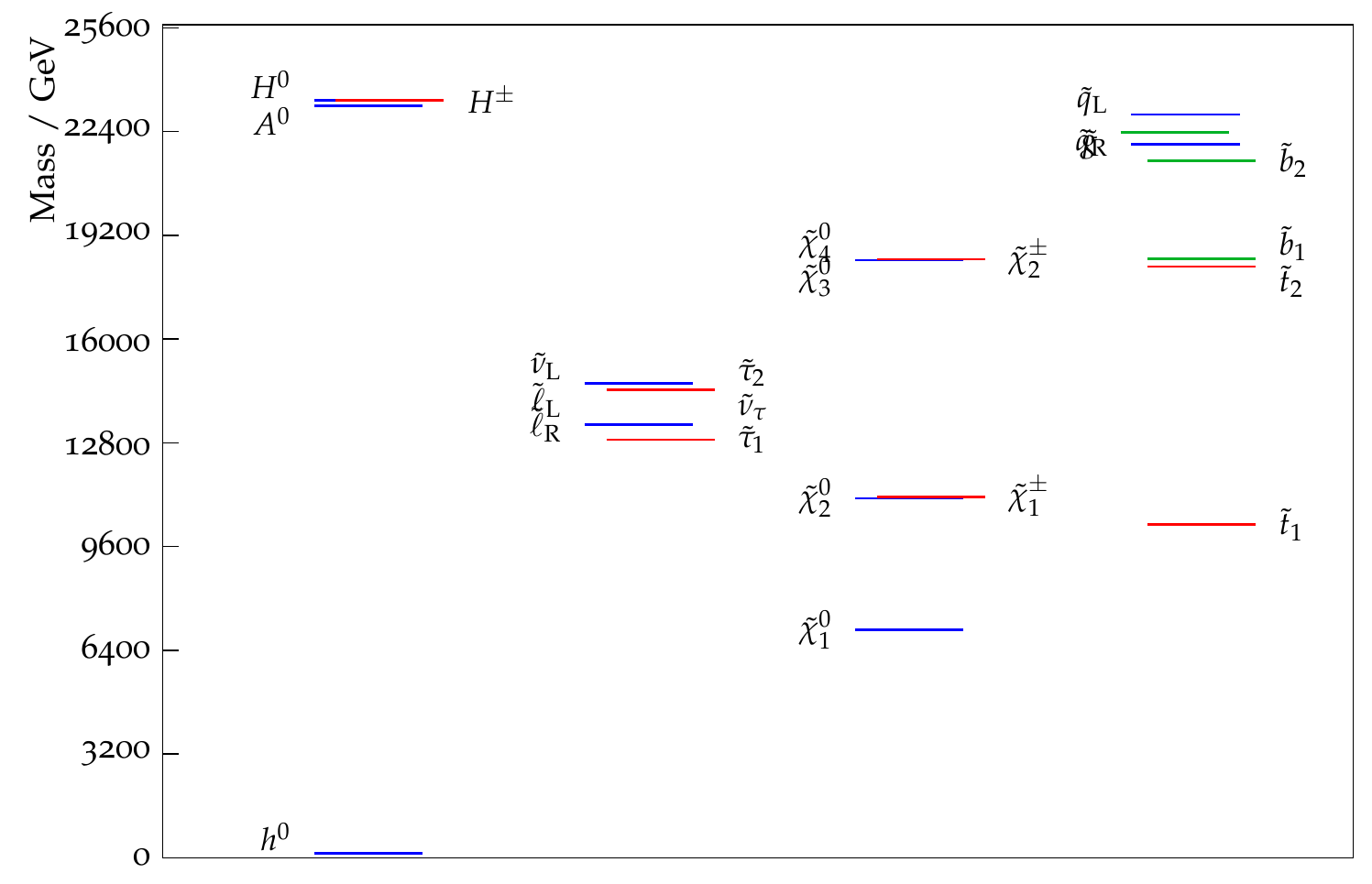} \includegraphics[width=8cm]{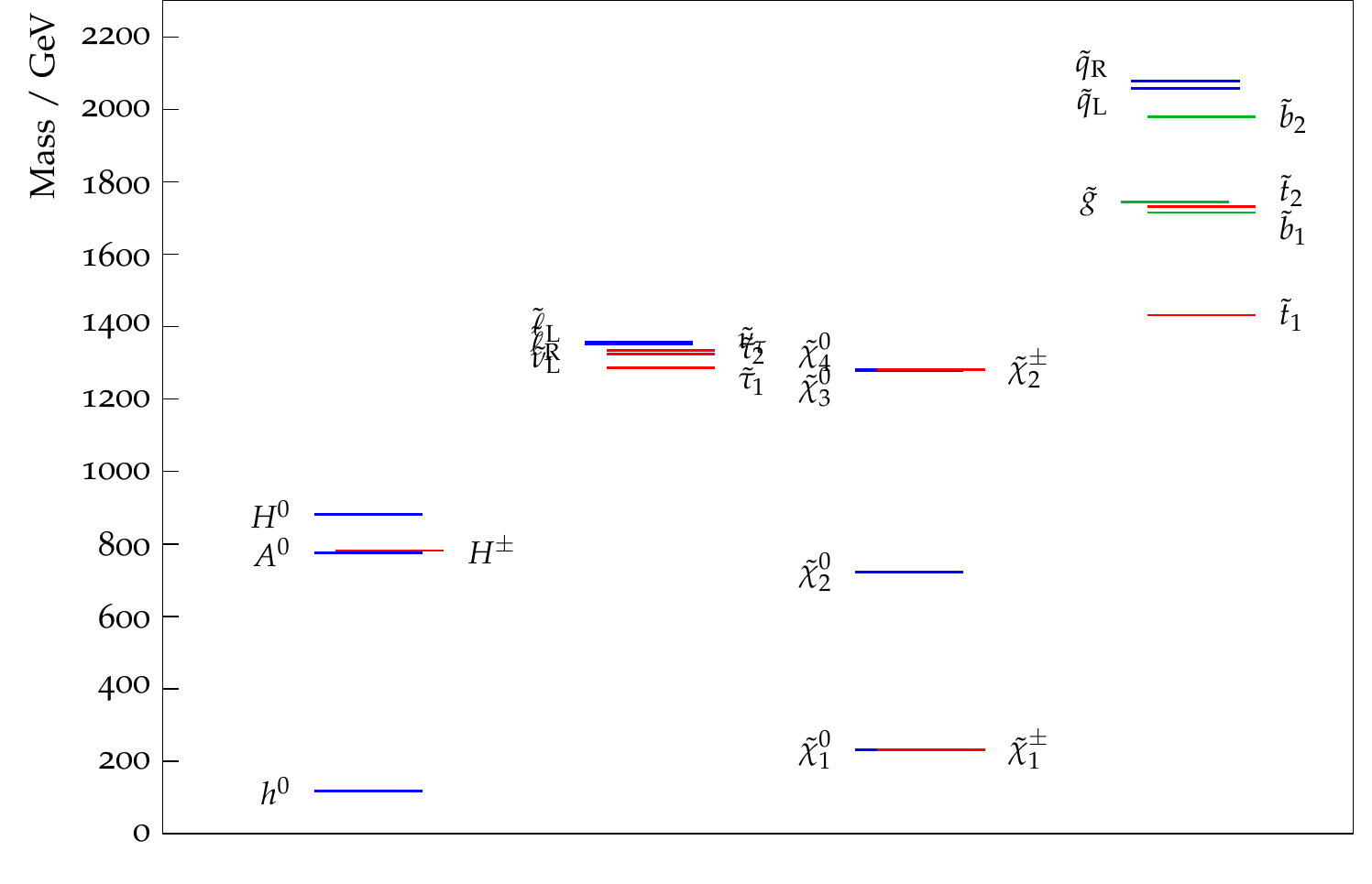}\\
\end{center}
\caption{\label{fig_spectra}Mass spectra of the six benchmark points. Note that the scales are not identical for all spectra.}
\end{figure}

We have seen in the previous section that the relic density imposes severe constraints to the parameter spaces, excluding a 
major part of the AMSB scenarios. However, the relic density calculation is generally based on the simplistic standard model of 
cosmology. In this section, we reinterpret the previous results by considering four different alternatives to the cosmological 
standard scenario. For this study, we focus our interest on six different benchmark points, which are described in Table \ref{points}
and are representative of the allowed parameter space in the different models. 
The mass spectra associated to these points are shown in Fig. \ref{fig_spectra}.

\begin{table}[!t]
\hspace*{0.4cm}
\begin{tabular}{|c|c|c|c|c|c|c|c|}
\hline
Point & Model & $\Omega_{DM} h^2$ & $m_0$ (GeV) & $\alpha$ & $m_{3/2}$ (TeV) & $\tan\beta$ & $M_A$(GeV)\\[1mm]
\hline\hline
A & mAMSB & $3.33 \times 10^{-4}$ & 1000 & n/a & 80 & 30 &  1060.5\\[1mm]
\hline
B & mAMSB & $4.63 \times 10^{-10}$ & 2000 & n/a & 20 & 40 & 1322.8\\[1mm]
\hline
C & HCAMSB & $3.24 \times 10^{-4}$ & n/a & 0.1 & 80 & 10 & 1931.3\\[1mm]
\hline
D & MMAMSB & 5.98 & n/a & 10 & 20 & 30 & 1904.4\\[1mm]
\hline
E & MMAMSB & $6.95 \times 10^{2}$ & n/a & 20 & 100 & 10 & 2320.5\\[1mm]
\hline
F & mNAMSB & $1.21 \times 10^{-2}$ & 1300 & n/a & 70 & 20 & 770\\[1mm]
\hline
\end{tabular}
\caption{Benchmark points for testing alternative cosmology scenarios. All these points have $\mu > 0$ and are in agreement with 
all the flavour and direct search constraints but would be excluded by WMAP constraints based on the standard cosmology. 
For the point F extra parameters are needed to specify a point in the parameter space, which are chosen to be: $\lambda=-0.1$,
$\kappa=0.5$ and $A_\kappa= 1500$ GeV. 
\label{points}}
\end{table}

\subsection{Alternative cosmological scenarios}

In the following, we consider that dark matter is composed of exclusively one particle produced thermally.\\
The density number of supersymmetric particles is determined by the Boltzmann equation \cite{relic_calculation}:
\begin{equation}
\frac{dn}{dt} = - 3 H n - \langle \sigma v \rangle (n^2 - n^2_{eq})\;,\label{boltzmann}
\end{equation}
where $n$ is the number density of supersymmetric particles, $\langle \sigma v \rangle$ is the thermally averaged annihilation 
cross-section, $H$ is the Hubble parameter, $n_{eq}$ is the relic particle equilibrium number density. The expansion rate $H$ is 
determined by the Friedmann equation:
\begin{equation}
 H^2=\frac{8 \pi G}{3}\rho_{rad} \;,\label{friedmann}
\end{equation}
where $\rho$ is the total energy density of the Universe. The entropy evolution reads:
\begin{equation}
\frac{ds}{dt} = - 3 H s\label{entropy_evolution} \;,
\end{equation}
where $s$ is the total entropy density. Solving and evolving simultaneously Eqs. (\ref{boltzmann}), (\ref{friedmann}) and 
(\ref{entropy_evolution}) enable to compute the relic density in our present Universe. In the standard cosmology, the dominant 
component before Big-Bang Nucleosynthesis is considered to be radiation, which is constituted of all relativistic particles. This 
assumption is however relaxed in many cosmological scenarios. The last two equations can indeed be written more generally as \cite{Arbey:2009gt}
\begin{eqnarray}
H^2 &= &\frac{8 \pi G}{3} (\rho_{rad} + \rho_D)  \;,\label{friedmannmod}\\
\frac{ds}{dt} &=& - 3 H s + \Sigma_D \label{entropy_evolutionmod} \;,
\end{eqnarray}
$\rho_D$ parametrises a modified evolution of the total density of the Universe, beyond radiation density $\rho_{rad}$. 
$\Sigma_D$ parametrises here effective entropy fluctuations due to unknown properties of the Early Universe. The radiation 
energy and entropy density evolutions are known and can be written as usual:
\begin{equation}
\rho_{rad}=g_{\rm{eff}}(T) \frac{\pi^2}{30} T^4 \;, \qquad\qquad s_{rad} = h_{\rm{eff}}(T) \frac{2\pi^2}{45} T^3 \;. \label{srad}
\end{equation}
In the following, we consider two scenarios in which the energy density is modified, and two scenarios with a modified entropy 
content.

\subsection{Quintessence}
The quintessence model is one of the most well-known models for dark energy. It is based on a cosmological scalar field which has 
presently a negative constant pressure $P$ and a positive constant energy density $\rho$ such as $P \approx - \rho$ 
\cite{quintessence2}. This behaviour is achieved when the kinetic term of the scalar field equilibrates the potential. However, in the 
early Universe, the scalar field has a dominating kinetic term, leading to a positive pressure such as $P \approx \rho$. During this 
period, the energy density was varying very quickly, such as $\rho \propto T^6$. We study here a quintessence scenario in which 
the quintessence field before Big-Bang Nucleosynthesis was dominating the expansion of the Universe. In this case \cite{Arbey:2008kv}:
\begin{equation}
 \rho_D(T) = \kappa_\rho \rho_{rad}(T_{BBN}) \left(\frac{T}{T_{BBN}}\right)^6 \;,
\end{equation}
where $\kappa_\rho$ is the proportion of quintessence to radiation at the BBN temperature ($\sim$1 MeV). We consider that 
$\kappa_\rho$ is a free parameter, which can be constrained using the BBN abundance constraints. To compute the abundance 
of the elements produced during the primordial nucleosynthesis, we use the code AlterBBN  \cite{alterbbn} integrated into SuperIso Relic. Comparing the abundances to the observational constraints, we obtain limits on $\kappa_\rho$.

\subsection{Late decaying inflaton}
The second scenario we consider here is a late decay of an inflaton field. The inflaton field is a scalar field which is considered to 
be responsible for the rapid inflation of the early Universe. Generally, the inflaton is considered to decay into standard particles 
much before the relic decoupling from the primordial soup. However, several models evoke the possibility of a late decay of the 
inflaton, around the time of BBN. From \cite{Gelmini:2006pw, Gelmini:2006pq}, there exist cosmological models in 
which the late decay of a scalar field reheats the Universe to a low reheating temperature , which can be smaller than 
the freeze-out temperature, without spoiling primordial nucleosynthesis. The decay of this scalar field into radiation increases the 
entropy and modifies the expansion rate of the Universe. We consider here such a scenario in which we neglect the eventual 
entropy production, and we takes \cite{Arbey:2008kv}
\begin{equation}
 \rho_D(T) = \kappa_\rho \rho_{rad}(T_{BBN}) \left(\frac{T}{T_{BBN}}\right)^8 \;.
\end{equation}
The exponent is here increased from 6 to 8 in comparison to the quintessence field, as mentioned also in 
\cite{reheating5,reheating6}. Such a modification of the expansion rate can be also achieved in a Universe with 
extra-dimensions modifying the Friedmann equations \cite{Okada:2004nc}.

\subsection{Primordial entropy production}
In this third scenario, we assume that a primordial entropy production due to an unknown component occurs. In general, such an 
entropy production should be accompanied with energy production, but to better estimate the deviation of the relic density in such 
a cosmological scenario, we neglect here the energy production, and we consider that the Universe has, in addition to the 
standard radiation entropy density, a dark entropy density evolving like \cite{Arbey:2009gt}
\begin{equation}
 s_D(T) = \kappa_s s_{rad}(T_{BBN}) \left(\frac{T}{T_{BBN}}\right)^3 \;,
\end{equation}
where $\kappa_s $ is the ratio of effective dark entropy density over radiation entropy density at the time of BBN.
The corresponding entropy production is related to $s_D$ by the relation
\begin{equation}
\Sigma_D = \sqrt{\frac{4 \pi^3 G}{5}} \sqrt{1 + \tilde{\rho}_D} T^2 \left[\sqrt{g_{eff}} s_D - \frac13  \frac{h_{eff}}{g_*^{1/2}} T \frac{ds_D}{dT}\right] \;, \label{SigmaD}
\end{equation}
with
\begin{equation}
g_*^{1/2} = \frac{h_{eff}}{\sqrt{g_{eff}}}\left(1+\frac{T}{3 h_{eff}} \frac{dh_{eff}}{dT}\right) \;,
\end{equation}
and
\begin{equation}
\tilde{\rho}_D=\frac{\rho_{D}}{\rho_{rad}} \;.
\end{equation}

\subsection{Late reheating}
In this last scenario, the reheating temperature $T_{RH}$ is smaller than the neutralino 
freeze out temperature ($T_{f.o.} \simeq m_{\chi}/20$ GeV) \cite{Gelmini:2006pq}, $T_{RH}$ should be considered as a cosmological parameter that 
can take any value around a few MeV and then the neutralinos decoupled from the plasma before the end of the reheating process 
so their relic density will differ from its standard value. We consider that the inflaton decays around the time of Big-Bang 
nucleosynthesis, generating entropy. From the standard late reheating scenarios, we assume that the entropy production evolves like \cite{Arbey:2009gt}
\begin{equation}
\Sigma_D(T) = \kappa_\Sigma \Sigma_{rad}(T_{BBN}) \left(\frac{T_{BBN}}{T}\right)
\end{equation}
for $T > 1$ MeV and that this entropy production stops at the time of BBN. We again neglect the energy production or non-thermal 
production of particles in order to better understand the effects of a reheating entropy production on the relic density. In term of 
entropy density, we have
\begin{equation}
 s_D(T) = 3 \sqrt{\frac{5}{4 \pi^3 G}} h_{eff} T^3 \int_0^T dT' \frac{g_*^{1/2}\Sigma_D(T')}{\sqrt{1+\dfrac{\rho_D}
 {\rho_{rad}}}h_{eff}^2 T'^6}\;.
\end{equation}

\subsection{BBN constraints and modified relic density}
The different scenarios do not have impact on the cosmological observations, but they can modify the abundance of the 
elements. We compute with AlterBBN \cite{alterbbn} the abundance of the elements in the different scenarios for each benchmark 
points, varying the single parameter for a given scenario, and we apply the following conservative constraints \cite{jedamzik06}:
\begin{eqnarray}
0.240 < Y_p < 0.258\;, \qquad 1.2 \times 10^{-5} < \!~^2\!H/H < 5.3 \times 10^{-5}\;,\qquad\qquad\label{BBNconstraints}\\
\nonumber 0.57 < \!~^3\!H/\!~^2\!H < 1.52\;, \qquad  \!~^7\!Li/H > 0.85 \times 10^{-10}\;,\qquad \!~^6\!Li/\!~^7\!Li < 0.66\;,
\end{eqnarray}
for the helium abundance $Y_p$ and the primordial $^2\!H/H$, $^3\!H/\!~^2\!H$, $^7\!Li/H$ and $^6\!Li/\!~^7\!Li$ ratios.

\begin{figure}[p!]
\begin{center}
\includegraphics[width=4.3cm]{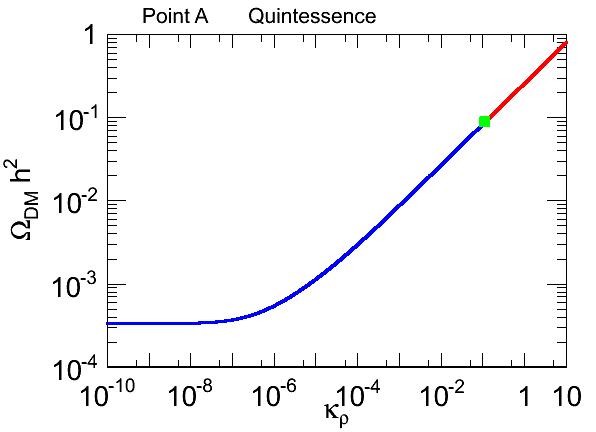}\includegraphics[width=4.3cm]{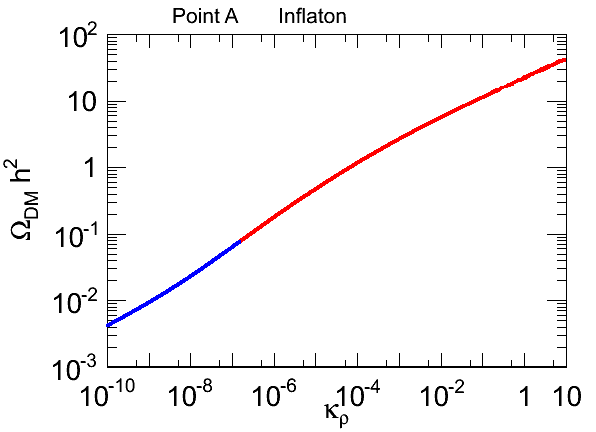}\includegraphics[width=4.3cm]{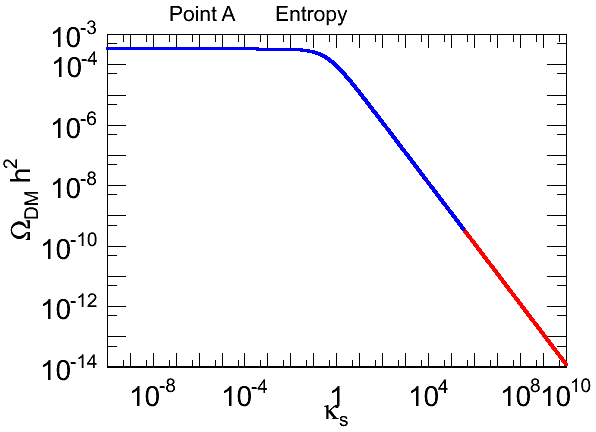}\includegraphics[width=4.3cm]{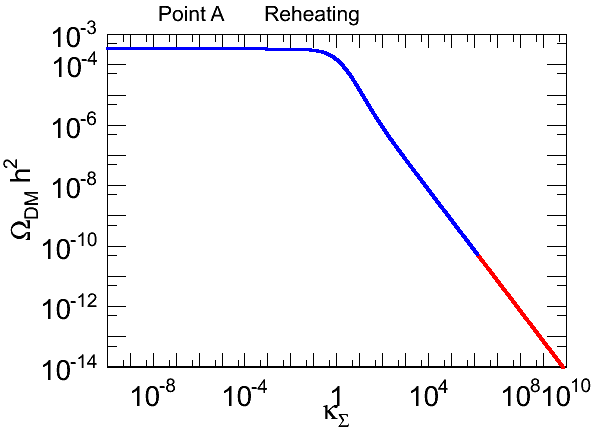}\\
\includegraphics[width=4.3cm]{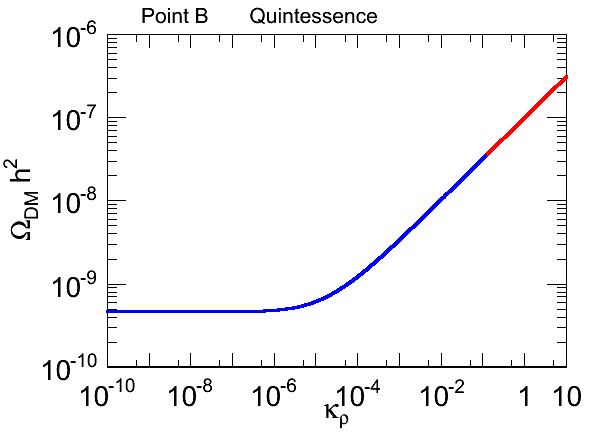}\includegraphics[width=4.3cm]{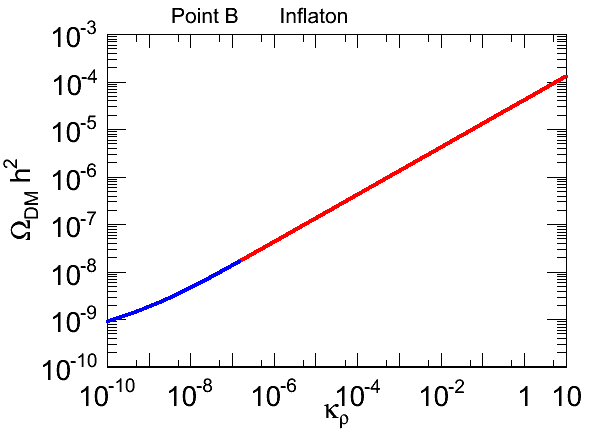}\includegraphics[width=4.3cm]{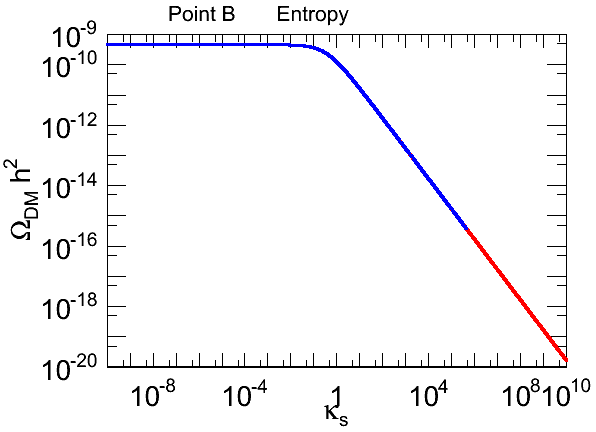}\includegraphics[width=4.3cm]{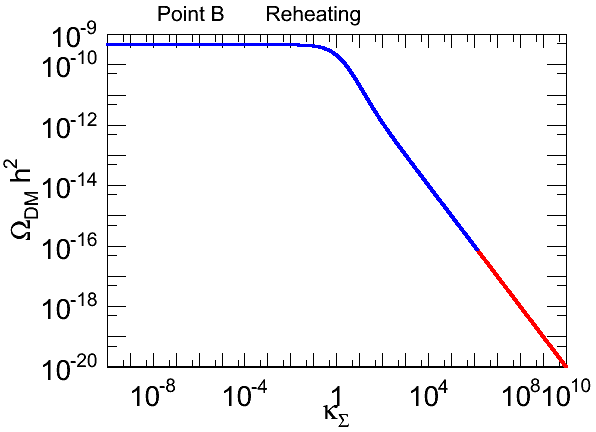}\\
\includegraphics[width=4.3cm]{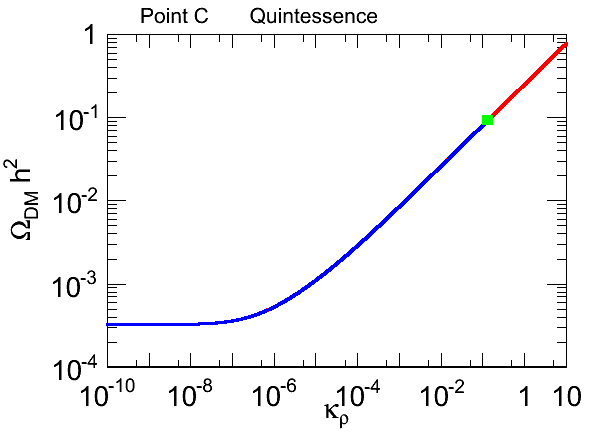}\includegraphics[width=4.3cm]{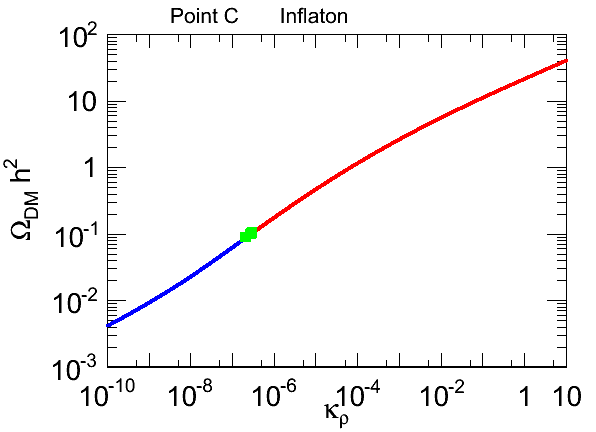}\includegraphics[width=4.3cm]{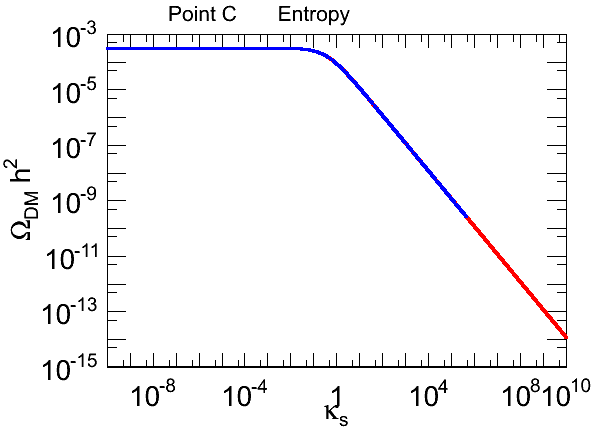}\includegraphics[width=4.3cm]{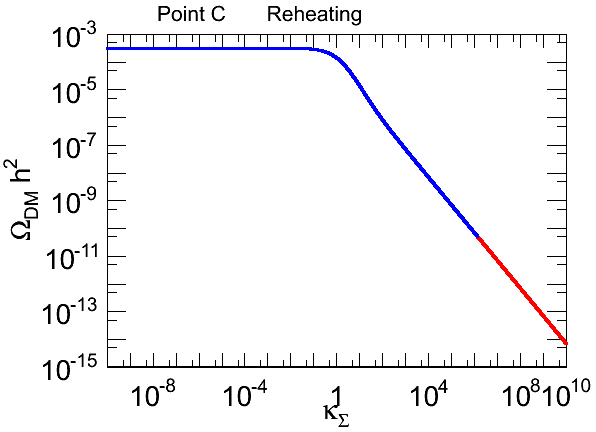}\\
\includegraphics[width=4.3cm]{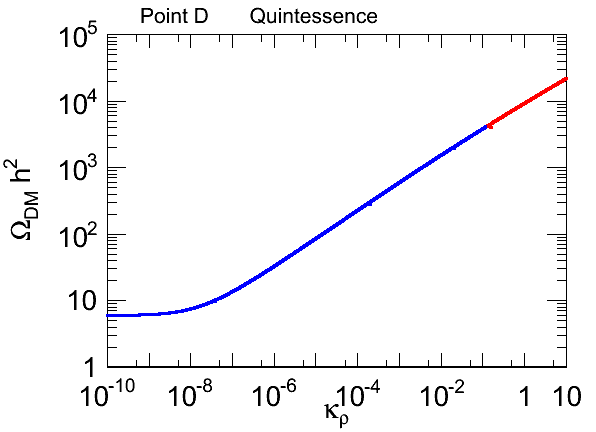}\includegraphics[width=4.3cm]{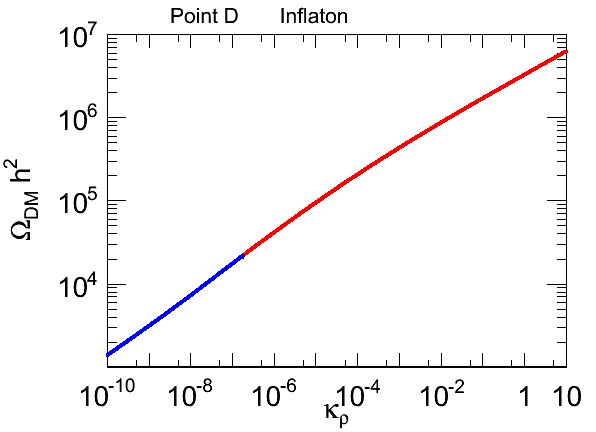}\includegraphics[width=4.3cm]{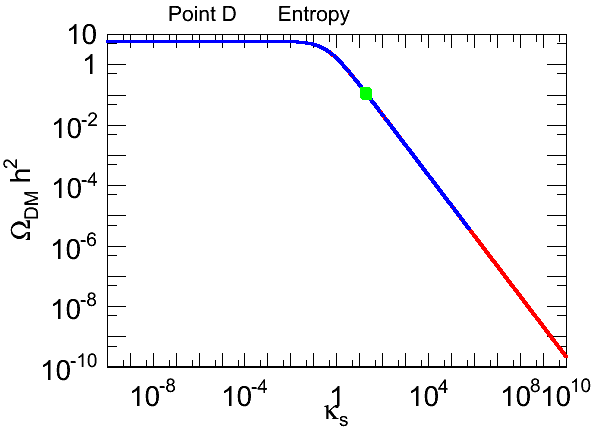}\includegraphics[width=4.3cm]{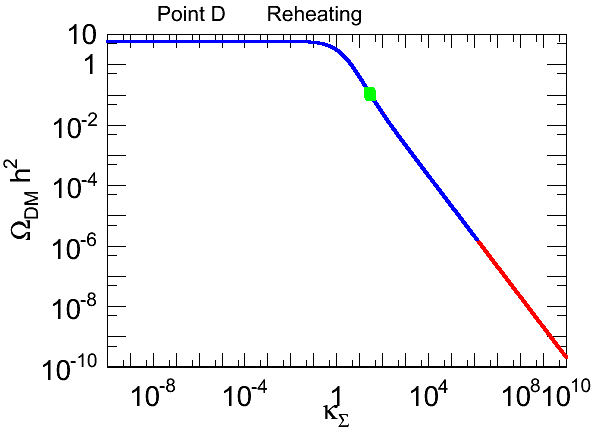}\\
\includegraphics[width=4.3cm]{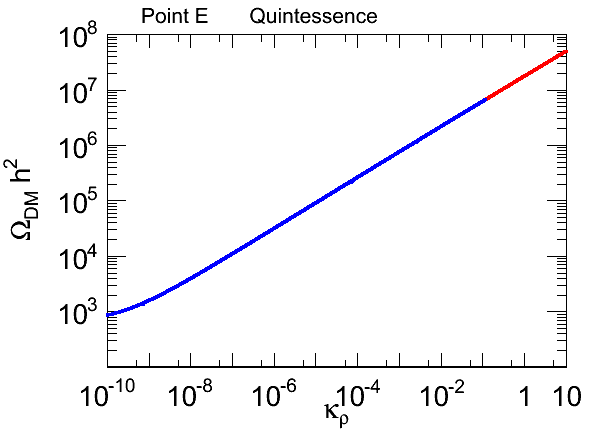}\includegraphics[width=4.3cm]{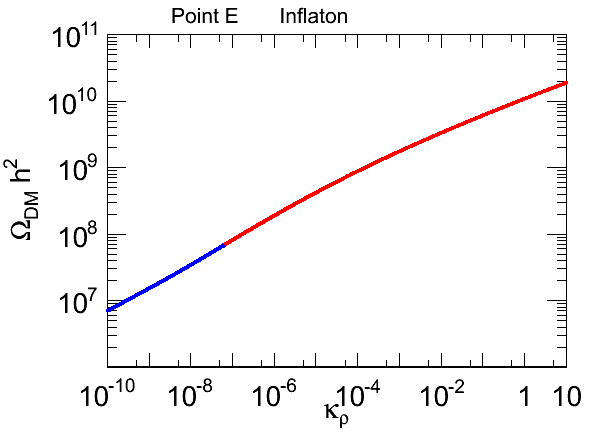}\includegraphics[width=4.3cm]{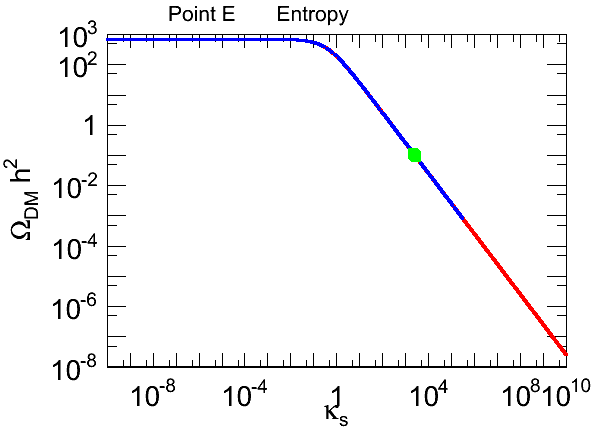}\includegraphics[width=4.3cm]{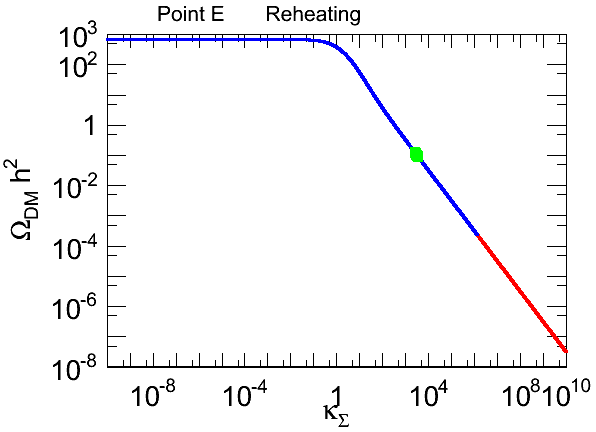}\\
\includegraphics[width=4.3cm]{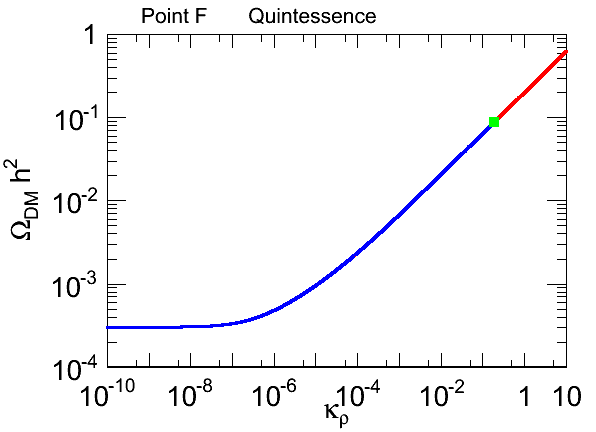}\includegraphics[width=4.3cm]{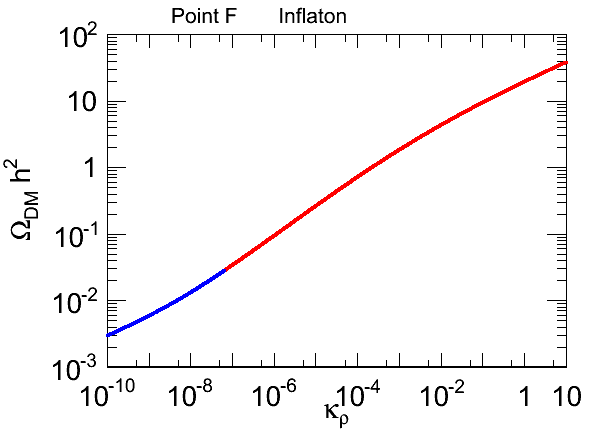}\includegraphics[width=4.3cm]{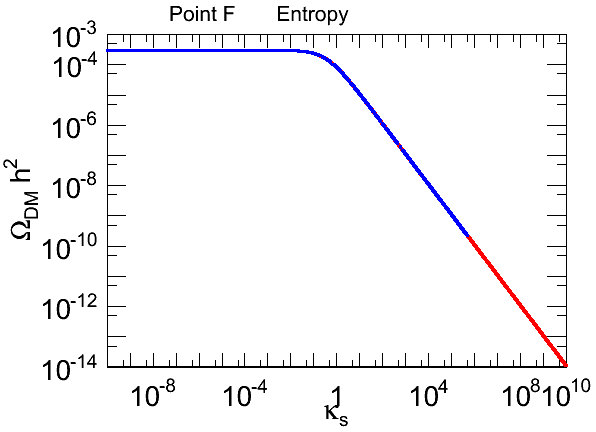}\includegraphics[width=4.3cm]{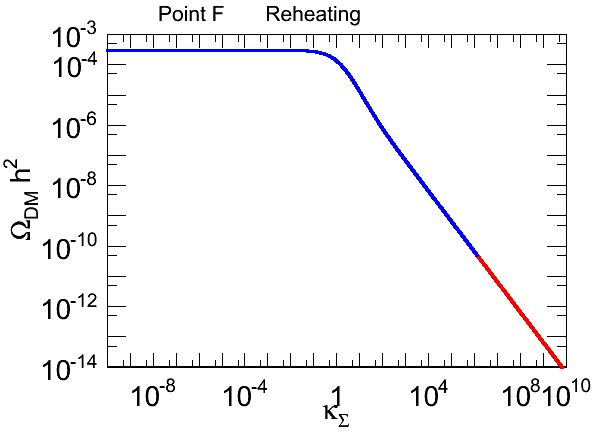}
\end{center}
\caption{\label{fig_bbn}Relic density in function of the cosmological model parameters. The red lines correspond to points excluded 
by the BBN constraints in Eq. (\ref{BBNconstraints}), whereas the blue lines correspond to a region with a correct abundance of the elements. The green squares correspond to points allowed by the BBN constraints and giving a relic density in agreement with the WMAP dark matter interval.}
\end{figure}

In Fig. \ref{fig_bbn}, we consider the relic density calculated for each of the benchmark points (A-F, from top to bottom) and for different cosmological scenarios (from left to right).

These plots reveal that the quintessence and inflaton scenarios globally increase the relic density, while the entropy and reheating scenarios decrease it. The comparison with the BBN constraints is also represented, and the red part of the curves is excluded at 95\% C.L., while the blue part gives a correct abundance of the elements. As general features, the quintessence and inflaton scenarios can increase the relic density by three orders of magnitudes without interfering with the BBN constraints, and the entropy and reheating scenarios can decreased to a factor of $10^6$. Therefore, apart from point B which has an extremely low relic density value in the standard cosmological scenario, all the other benchmark points can have a relic density value compatible with the cosmological observations if a non minimal cosmological scenario is considered.

\section{Generalised relic density constraints}
\label{sec:constr}
\begin{figure}[p!]
\begin{center}
\includegraphics[width=16cm]{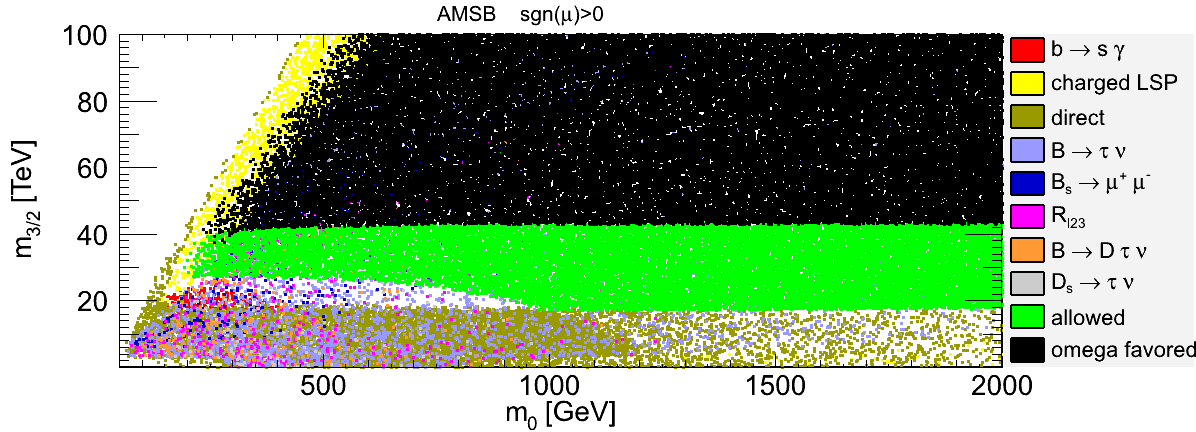}\\
\includegraphics[width=16cm]{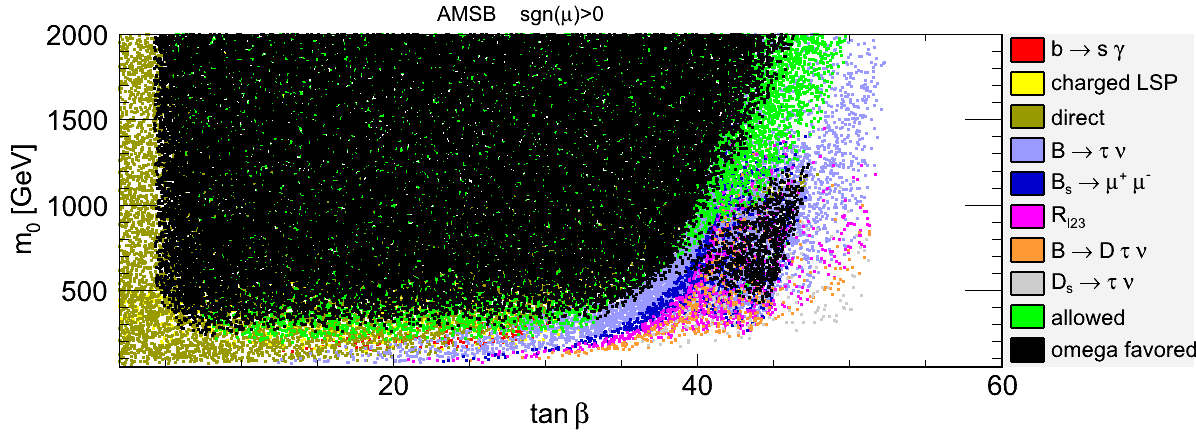}\\
\includegraphics[width=16cm]{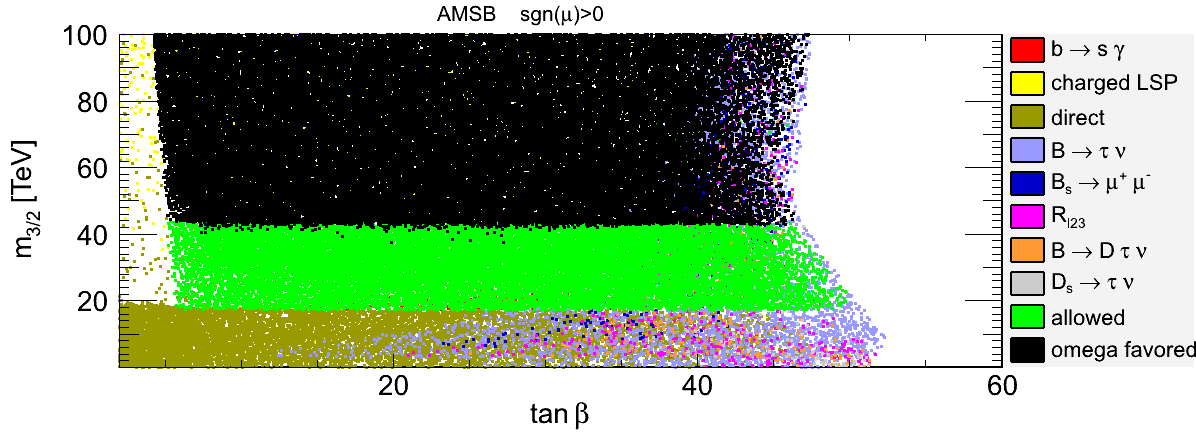}
\end{center}
\caption{\label{fig_amsb_reloaded}Constraints on the minimal AMSB parameter space. The exclusion regions are plotted in the 
order given in the legend. The red zones are excluded by the inclusive branching ratio of $B \to X_s \gamma$, the yellow ones 
correspond to charged LSP, the olive green area is excluded by direct collider constraints, the light blue zones are excluded by 
BR($B\to\tau \nu$), the dark blue zones by BR($B_s\to\mu^+ \mu^-$), the magenta zones by $R_{\ell 23}$, the orange zones by 
BR($B\to D \tau \nu$) and the grey zones by BR($D_s\to \tau \nu$). The green areas are in agreement with all the previously 
mentioned constraints. The black area corresponds to parameters in agreement with all constraints, including the revised relic 
density interval given in Eq. (\ref{newWMAP}).}
\end{figure}

\begin{figure}[p!]
\begin{center}
\includegraphics[width=16cm]{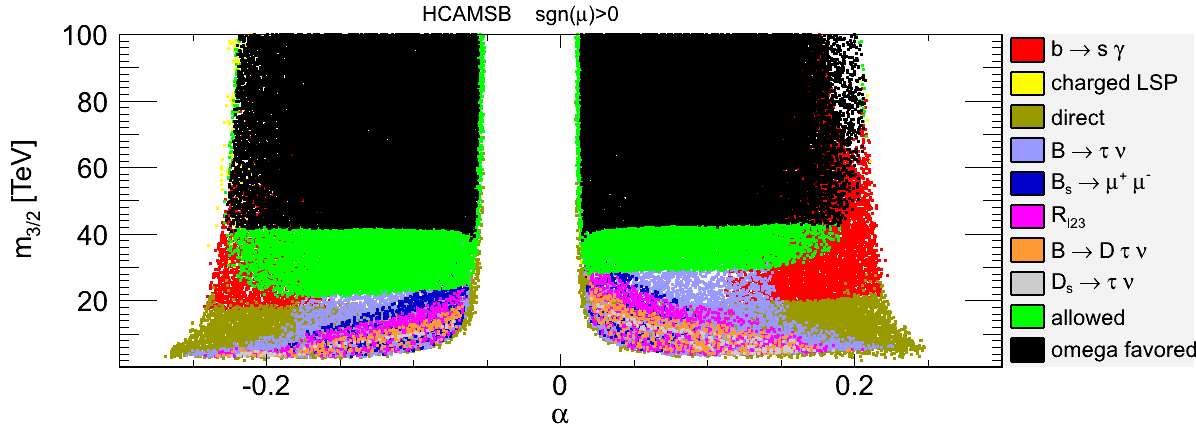}\\
\includegraphics[width=16cm]{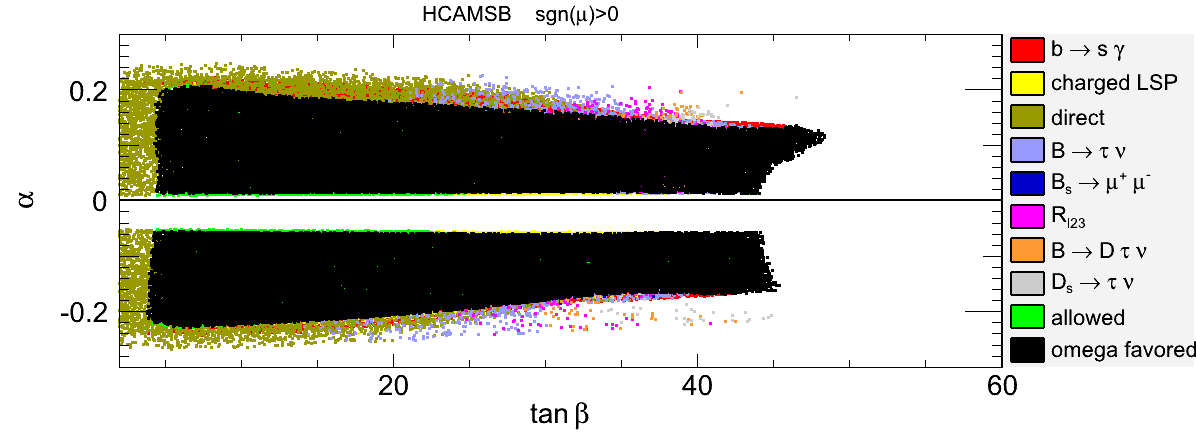}\\
\includegraphics[width=16cm]{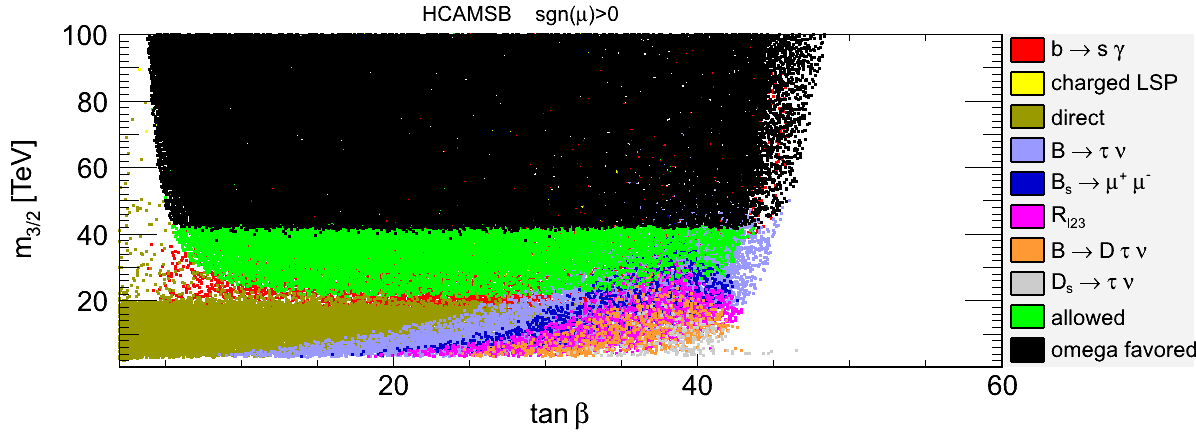}
\end{center}
\caption{\label{fig_hcamsb_reloaded}Constraints on the HCAMSB parameter space with the revised relic density interval given in Eq. (\ref{newWMAP}). 
The colour codes are the same as in Fig. \ref{fig_amsb_reloaded}.}
\end{figure}

\begin{figure}[p!]
\begin{center}
\includegraphics[width=16cm]{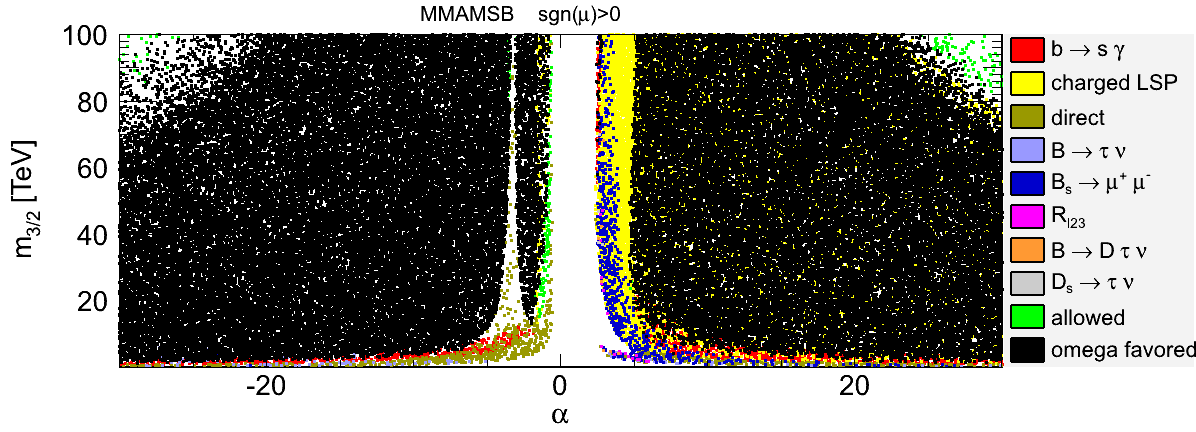}\\
\includegraphics[width=16cm]{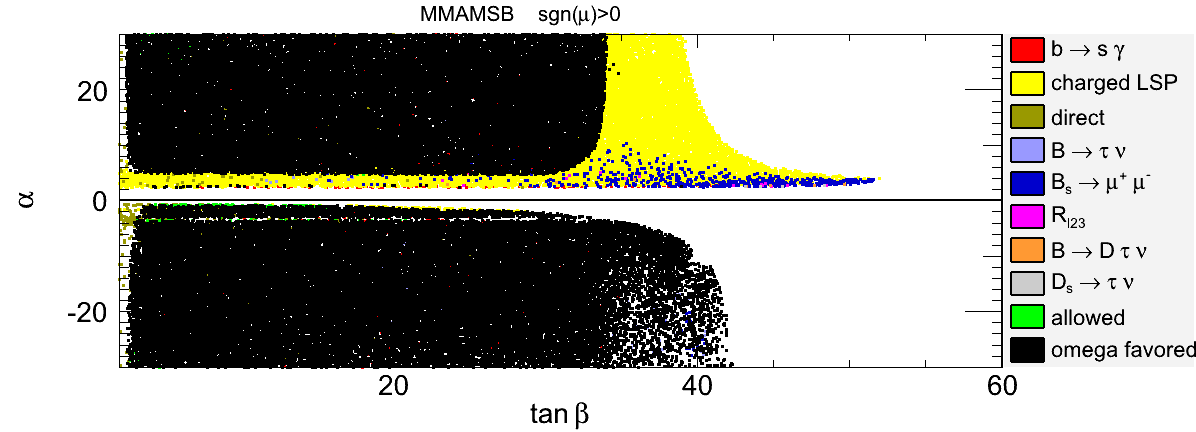}\\
\includegraphics[width=16cm]{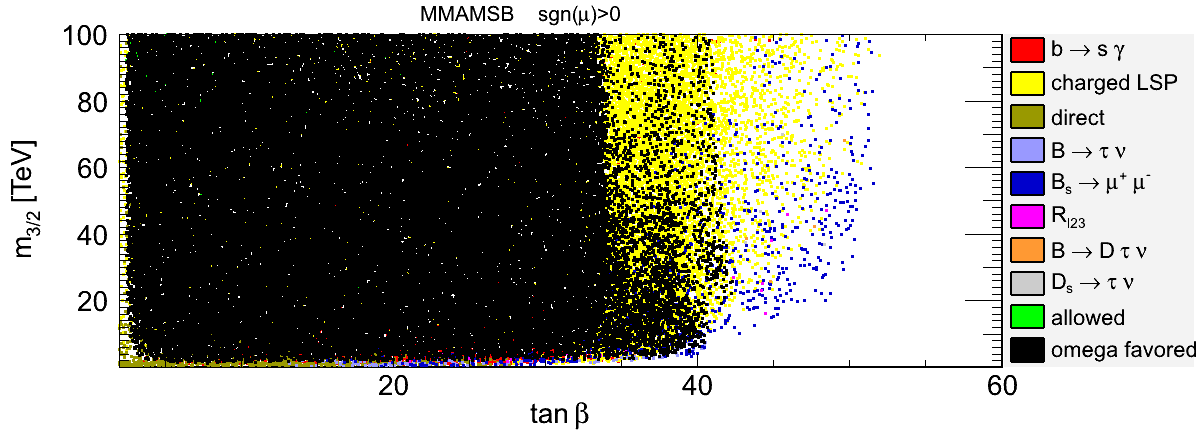}
\end{center}
\caption{\label{fig_mmamsb_reloaded}Constraints on the MMAMSB parameter space with the revised relic density interval given in Eq. (\ref{newWMAP}). The colour codes are the same as in Fig. \ref{fig_amsb_reloaded}.}
\end{figure}

\begin{figure}[p!]
\begin{center}
\includegraphics[width=16cm]{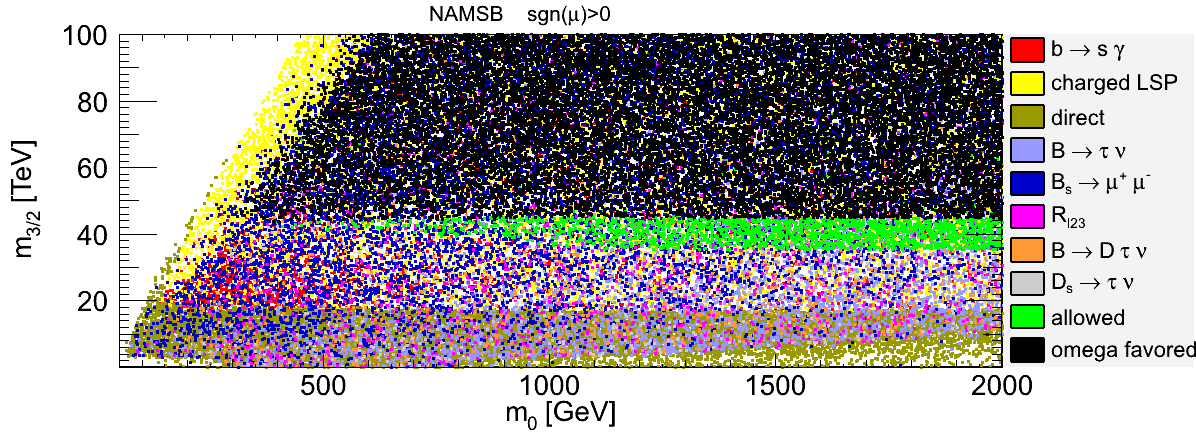}\\
\includegraphics[width=16cm]{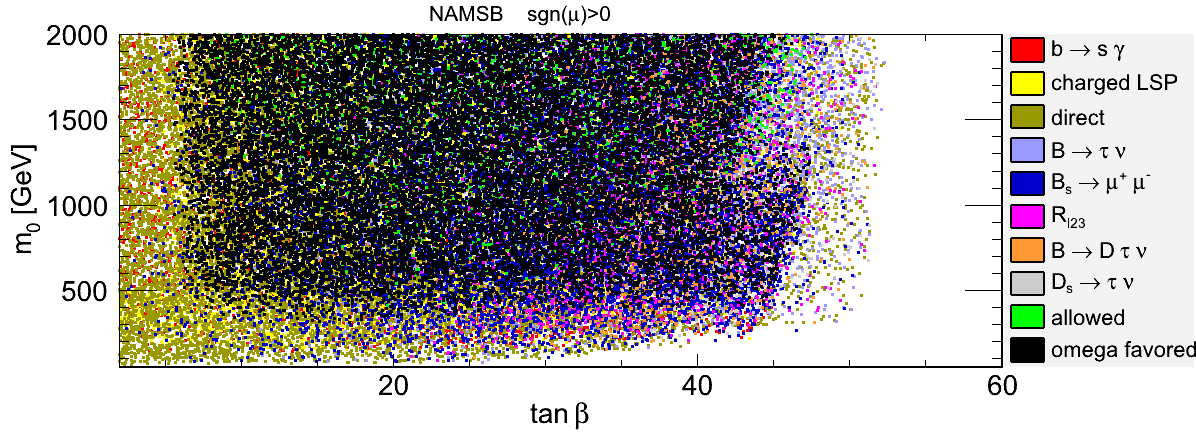}\\
\includegraphics[width=16cm]{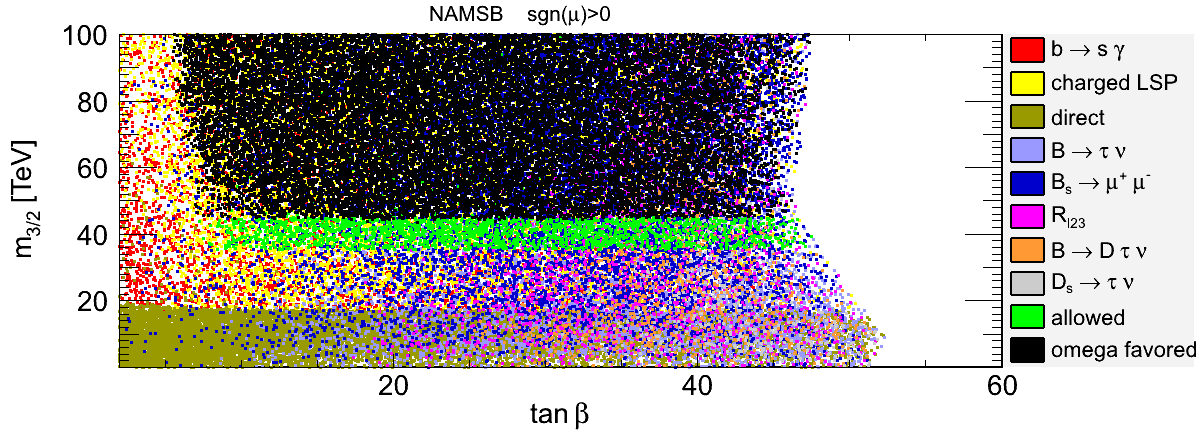}
\end{center}
\caption{\label{fig_namsb_reloaded}Constraints on the mNAMSB parameter space with the revised relic density interval given in Eq. (\ref{newWMAP}). 
The colour codes are the same as in Fig. \ref{fig_amsb_reloaded}.}
\end{figure}

\begin{figure}[p!]
\begin{center}
\includegraphics[width=16cm]{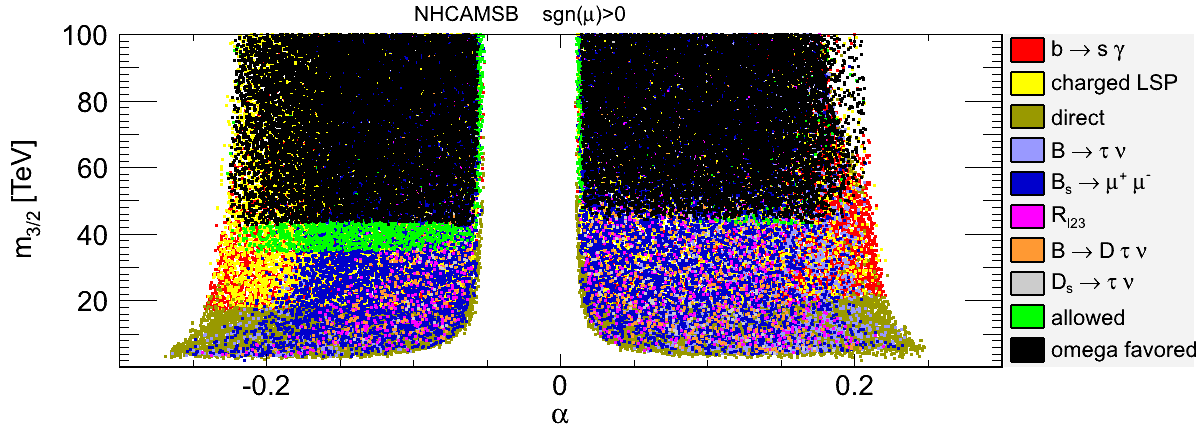}\\
\includegraphics[width=16cm]{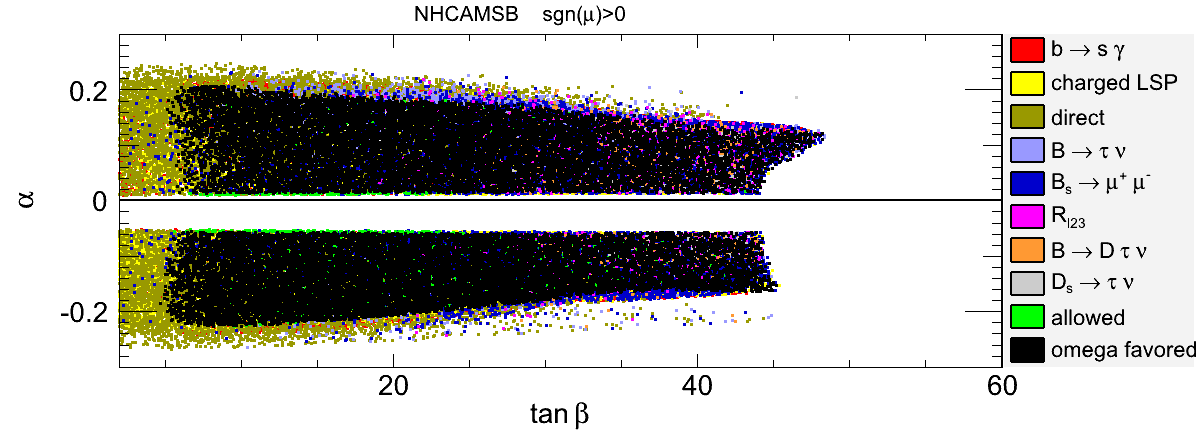}\\
\includegraphics[width=16cm]{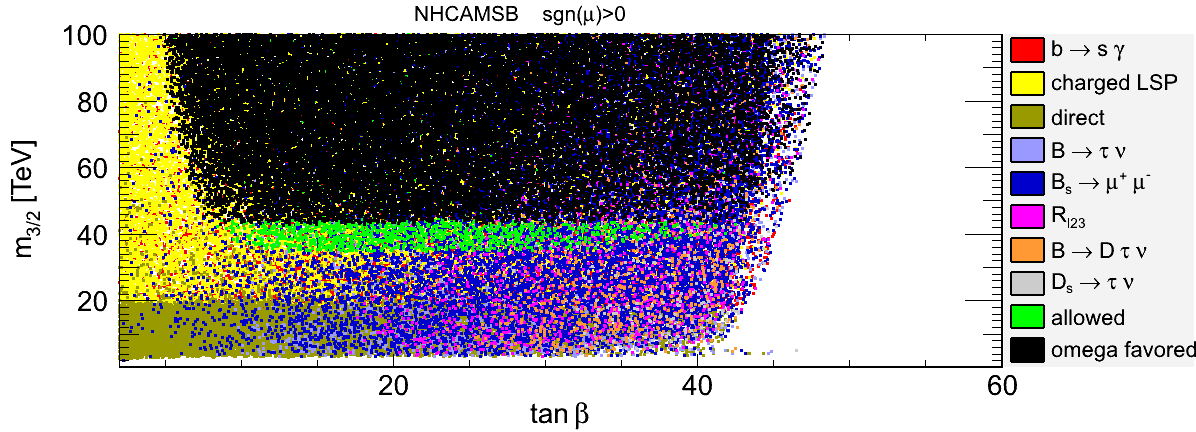}
\end{center}
\caption{\label{fig_nhcamsb_reloaded}Constraints on the NHCAMSB parameter space with the revised relic density interval given in Eq. (\ref{newWMAP}). 
The colour codes are the same as in Fig. \ref{fig_amsb_reloaded}.}
\end{figure}

\begin{figure}[p!]

\begin{center}
\includegraphics[width=16cm]{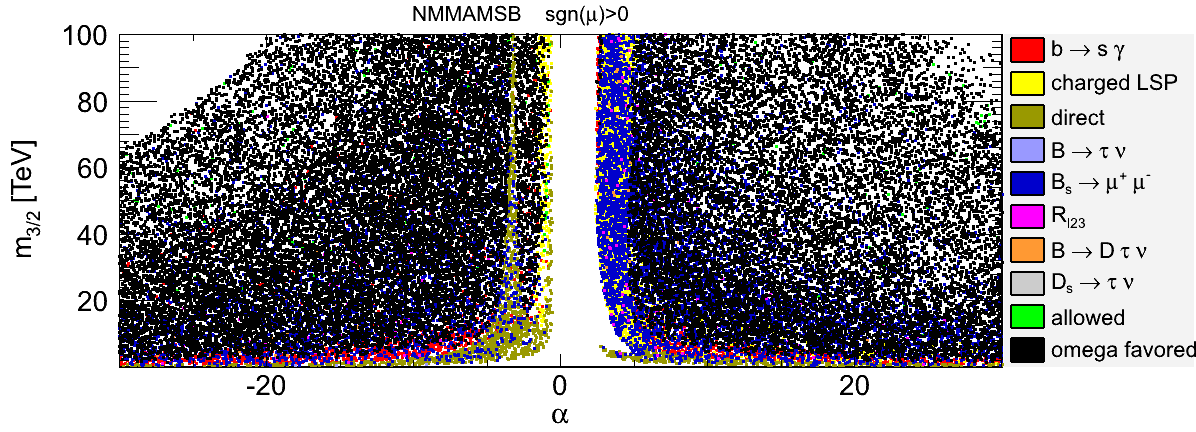}\\
\includegraphics[width=16cm]{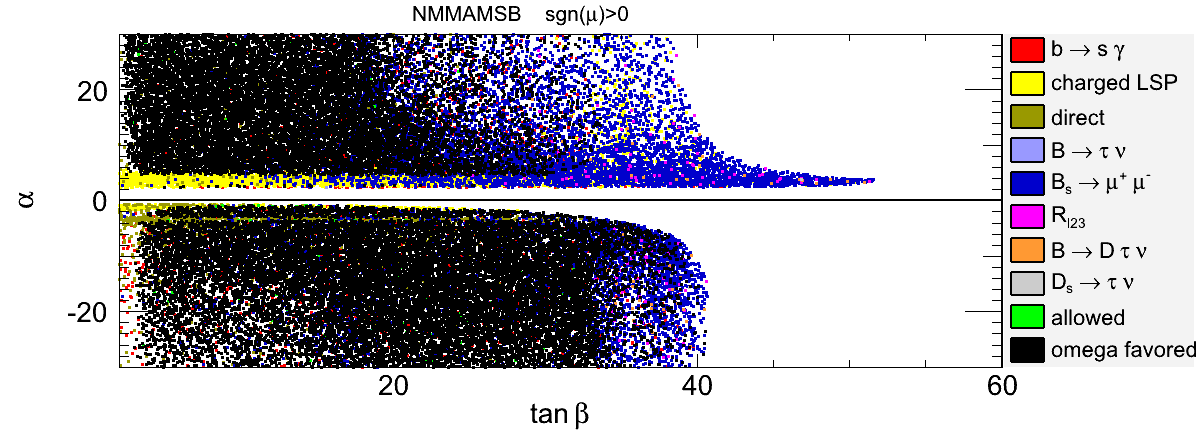}\\
\includegraphics[width=16cm]{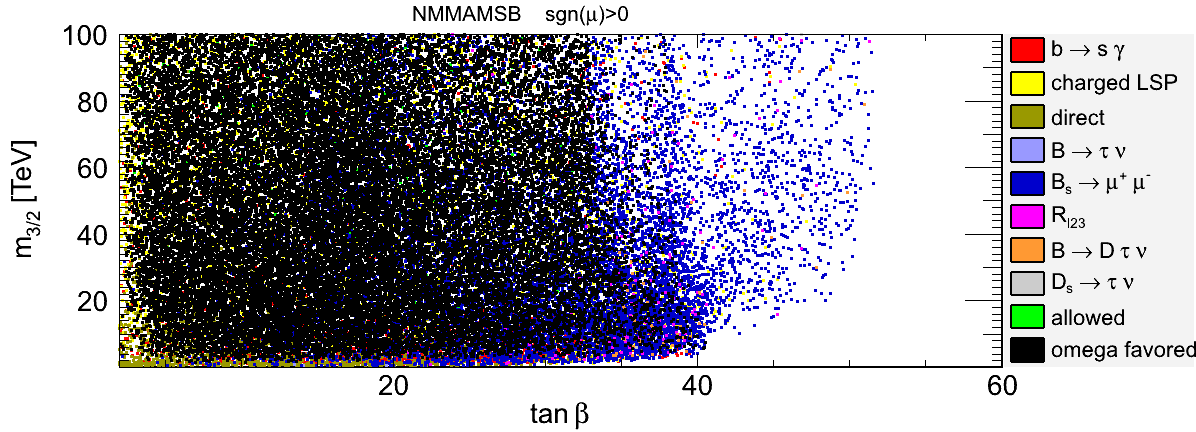}
\end{center}
\caption{\label{fig_nmmamsb_reloaded}Constraints on the NMMAMSB parameter space including the revised relic 
density interval. The colour codes are the same as in Fig. \ref{fig_amsb_reloaded}.}
\end{figure}

We have shown with different well-known cosmological scenarios, that the relic density constraints can be very strongly relaxed. Therefore, we propose to compare the relic density calculated in the standard model of cosmology to the following interval
\begin{equation}
10^{-4} < \Omega_{DM} h^2 < 10^5
\label{newWMAP}
\end{equation}
to take into consideration the fact that it is possible to increase any relic density calculated in the standard cosmology by three 
orders of magnitude, and to decrease it by six orders, with non-standard cosmological scenarios in agreement with the current cosmological data. We can re-apply the relic density constraints, and the results are shown 
in Figs. \ref{fig_amsb_reloaded}-\ref{fig_nmmamsb_reloaded}. As in the figures of section \ref{sec:models}, the green zones correspond to regions in agreement with all flavour and direct constraints, but not necessarily with the relic density constraint. We added in the figures black points, which correspond to regions also in agreement with the new dark matter interval. It is clear that the allowed regions are therefore much larger than with the initial relic density interval, but a surprising result is that even with the very large interval we use here for the relic density, the relic density constraint still excludes large part of the parameter spaces. In particular, in the mAMSB and HCAMSB scenarios and their NMSSM counterparts, the relic density constraints clearly exclude the region $m_{3/2} \lesssim 40$ TeV. The MMAMSB scenario however is not constrained anymore when using the new dark matter interval.



\section{LHC phenomenology}
\label{sec:lhc}
The benchmark points selected for testing alternative cosmology scenarios are in agreement with precision flavour and 
direct search constraints. It turns out that the phenomenology expected at the LHC is quite peculiar. The mass spectra for the 
benchmark points we considered show that the lightest neutralino is the LSP and the lightest chargino
is in most cases very close in mass to the neutralino (points A, B, C, F). This will give rise to peculiar signatures due to the very 
limited number of open channels for the decay modes.  We analyse in more detail in the following each of the 
six benchmark points previously selected. Production of charginos and neutralinos takes place at the LHC via cascade 
decays of squark and gluinos and via the direct production channels
\begin{equation}
pp \to  {\tilde{\chi}}_i {\tilde{\chi}}_j + X
\end{equation}
where the s-channel exchange of an off-shell W or Z or photon, and the contribution of SUSY-QCD diagrams are important. 
Indeed these cross-sections receive important SUSY-QCD corrections, typical values are given in \cite{Beenakker:1999xh} or can be 
obtained using a Monte Carlo program including the relevant K-factors or the detailed next-to-leading matrix elements. 

\subsection{mAMSB point A}
The minimal AMSB scenario is disfavoured by the standard cosmology as very far from the WMAP dark matter allowed interval. The 
region allowed in our more general analysis typically favours points in which the lightest chargino and neutralino are very close in
mass and not so heavy. For point A, $m_{{\tilde{\chi}}_1^0}=231.76$ GeV and $m_{{\tilde{\chi}}_1^+}=231.93$ GeV so that 
the mass splitting is only 170 MeV. Due to this small mass splitting, the open decay modes for the ${\tilde{\chi}}_1^+$ are
 (neglecting very small modes below the $10^{-3}$ level) ${\tilde{\chi}}_1^+ \to {\tilde{\chi}}_1^0 l \nu$, where 
\begin{eqnarray}
{\mathrm {BR}}({\tilde{\chi}}_1^+ \to {\tilde{\chi}}_1^0 \mu^+ \nu_\mu) &\simeq& 1.87 \times 10^{-2} \\
{\mathrm {BR}}({\tilde{\chi}}_1^+ \to {\tilde{\chi}}_1^0 e^+ \nu_e) &\simeq& 1.87 \times 10^{-2} \\
{\mathrm {BR}}({\tilde{\chi}}_1^+ \to   {\tilde{\chi}}_1^0 \pi^+ \to
 {\tilde{\chi}}_1^0 e^+ \nu_e ) &\simeq& 0.96 \;,
\end{eqnarray}
so that $l=e$ and $l=\mu $ have the same branching through the three  body decays while most of the signal 
with ${\tilde{\chi}}_1^0 e^+ \nu_e$ is through the two body production of a charged pion.
The next lightest particle is the ${\tilde{\chi}}_2^0$, decaying mainly to ${\tilde{\chi}}_1^+$ and W or ${\tilde{\chi}}_1^0$ and SM-like Higgs
as for the light Higgs for this point (118.15 GeV) the decay is kinematically allowed :
\begin{eqnarray}
{\mathrm {BR}}({\tilde{\chi}}_2^0 \to  {\tilde{\chi}}_1^\pm W^\mp ) &\simeq& 0.75 \\
{\mathrm {BR}}({\tilde{\chi}}_2^0 \to  {\tilde{\chi}}_1^0 h^0 ) &\simeq& 0.19\\
{\mathrm {BR}}({\tilde{\chi}}_2^0 \to  {\tilde{\chi}}_1^0 l^+ l^- ) &\simeq& 5.5 \times 10^{-3}\;,
\end{eqnarray}
where $l=e$, $\mu$.
Since the  branching of the mode ${\tilde{\chi}}_1^\pm \to {\tilde{\chi}}_1^0 l^\pm \nu$ is 100\% and the mode
${\tilde{\chi}}_2^0 \to  {\tilde{\chi}}_1^0 l^+ l^-$ is non-negligible one can study the clean trilepton signal usually
suggested at hadron colliders \cite{Baer:1992dc,Baer:1994nr}.

\subsection{mAMSB point B}
The benchmark point B is in the minimal AMSB scenario too. The situation is similar to the previous point with in this case 
the lightest chargino and neutralino lighter than the SM-like Higgs boson. For point B, $m_{{\tilde{\chi}}_1^0}=52.70$ GeV and 
$m_{{\tilde{\chi}}_1^+}=53.07$ GeV so that the mass splitting is only 63 MeV. The decay modes for the lightest chargino are 
similar to the previous case
\begin{eqnarray}
{\mathrm {BR}}({\tilde{\chi}}_1^+ \to {\tilde{\chi}}_1^0 \mu^+ \nu_\mu) &\simeq& 5.26 \times 10^{-2} \\
{\mathrm {BR}}({\tilde{\chi}}_1^+ \to {\tilde{\chi}}_1^0 e^+ \nu_e) &\simeq& 5.26 \times 10^{-2} \\
{\mathrm {BR}}({\tilde{\chi}}_1^+ \to   {\tilde{\chi}}_1^0 \pi^+ \to
 {\tilde{\chi}}_1^0 e^+ \nu_e ) &\simeq& 0.89\;,
\end{eqnarray}
and for the second neutralino
\begin{eqnarray}
{\mathrm {BR}}({\tilde{\chi}}_2^0 \to  {\tilde{\chi}}_1^\pm W^\mp ) &\simeq& 0.8 \\
{\mathrm {BR}}({\tilde{\chi}}_2^0 \to  {\tilde{\chi}}_1^0 h^0 ) &\simeq& 0.1\\
{\mathrm {BR}}({\tilde{\chi}}_2^0 \to  {\tilde{\chi}}_1^0 Z^0 ) &\simeq& 9.6 \times 10^{-2}\;.
\end{eqnarray}
The SM-Higgs boson (with a mass of 115.7 GeV) decays with a sizeable branching to pairs of lightest charginos and pairs of 
lightest neutralinos (branching $1.1 \times 10^{-1}$ and $7.9 \times 10^{-2}$ respectively) while the largest mode is 
$h^0 \to b{\bar b}$ with a branching of $6.5 \times 10^{-1}$.

\subsection{HCAMSB point C}
The HCAMSB scenario is disfavoured by standard cosmology and excluded by the WMAP dark matter allowed interval as 
for the minimal AMSB scenario discussed above. Also in this case the lightest chargino and neutralino are very close in
mass. The masses are $m_{{\tilde{\chi}}_1^0}=229.41$ GeV and $m_{{\tilde{\chi}}_1^+}=229.58$ GeV with a splitting of only
17 MeV. In this situation the lightest chargino decays are very close to the numbers for point A. The decay modes for the 
lightest chargino are : 
\begin{eqnarray}
{\mathrm {BR}}({\tilde{\chi}}_1^+ \to {\tilde{\chi}}_1^0 \mu^+ \nu_\mu) &\simeq& 1.72 \times 10^{-2} \\
{\mathrm {BR}}({\tilde{\chi}}_1^+ \to {\tilde{\chi}}_1^0 e^+ \nu_e) &\simeq& 1.72 \times 10^{-2} \\
{\mathrm {BR}}({\tilde{\chi}}_1^+ \to   {\tilde{\chi}}_1^0 \pi^+ \to
 {\tilde{\chi}}_1^0 e^+ \nu_e ) &\simeq& 0.96\;,
\end{eqnarray}
and for the second neutralino
\begin{eqnarray}
{\mathrm {BR}}({\tilde{\chi}}_2^0 \to  {\tilde{\chi}}_1^\pm W^\mp ) &\simeq& 0.67 \\
{\mathrm {BR}}({\tilde{\chi}}_2^0 \to  {\tilde{\chi}}_1^0 h^0 ) &\simeq& 7.3 \times 10^{-2}\\
{\mathrm {BR}}({\tilde{\chi}}_2^0 \to  {\tilde{\chi}}_1^0 Z^0 ) &\simeq& 0.26
\end{eqnarray}
In this case the trilepton mode $pp \to {\tilde{\chi}}_1^\pm {\tilde{\chi}}_2^0 \to 3 l + {\sla{E}}_T$ is especially interesting
as the branching fraction ${\tilde{\chi}}_1^+ \to {\tilde{\chi}}_1^0 l \nu$ is 100\% and the one for 
${\tilde{\chi}}_2^0 \to  {\tilde{\chi}}_1^0 Z^0$ is 26\%.

\subsection{MMAMSB point D}
The Mixed Modulus AMSB supersymmetry breaking scenario allows for viable dark matter candidates. Benchmark point D has the neutralino LSP with a mass $m_{{\tilde{\chi}}_1^0}=736$ GeV. The next-to-lightest
supersymmetric particle is the stau with a mass $m_{{\tilde{\tau}}}=860$ GeV while the lightest chargino and the second lightest neutralino are heavier and almost degenerate with a mass of 1095 and 1098 GeV respectively. The stau decays to
tau and the lightest neutralino 
\begin{equation}
{\mathrm {BR}}({\tilde{\tau}} \to {\tilde{\chi}}_1^0 \tau ) = 1\;,
\end{equation}
while the lightest chargino decays  mainly to
\begin{eqnarray}
{\mathrm {BR}}({\tilde{\chi}}_1^+ \to {\tilde{\tau}} \nu_\tau ) &\simeq& 0.84 \\
{\mathrm {BR}}({\tilde{\chi}}_1^+ \to  {\tilde{\chi}}_1^0 W ) &\simeq& 0.16\;.
\end{eqnarray}
The second lightest neutralino decays to stau and tau while as the SM-like Higgs is light enough 
to be produced (124.3 GeV), the decay $ {\tilde{\chi}}_2^0\to {\tilde{\chi}}_1^0 h^0$ is also kinematically open
\begin{eqnarray}
{\mathrm {BR}}({\tilde{\chi}}_2^0 \to {\tilde{\tau}} \tau ) &\simeq& 0.83 \\
{\mathrm {BR}}({\tilde{\chi}}_2^0 \to  {\tilde{\chi}}_1^0 h^0 ) &\simeq& 0.16\;.
\end{eqnarray}

\subsection{MMAMSB point E}
For the same model as in the previous paragraph we have also an example of a higher mass spectrum in which all supersymmetric partners are quite heavy with the neutralino LSP above 7 TeV and all other particles above 10 TeV.

\subsection{mNAMSB point F}
NMSSM with minimal AMSB scenario is one of the models considered in the previous sections. It is disfavoured by the relic 
density constraints if standard cosmology is assumed. Point F is in the allowed zone for flavour, low energy and direct 
search constraints and is allowed if standard cosmology constraints are relaxed as discussed in the previous sections. 
From the point of view of LHC searches, this point is similar to point A. For point F, $m_{{\tilde{\chi}}_1^0}=230.03$ GeV and 
$m_{{\tilde{\chi}}_1^+}=231.02$ GeV so that the mass splitting is 99 MeV. 
Apart from these two particles, the next 
supersymmetric particle (in terms of mass) is the second neutralino (see Fig. \ref{fig_spectra}) with a mass of 721 GeV.

\section{Conclusions and perspectives}
\label{sec:conclusion}
We have considered in details the constraints on different possible realisations of superconformal anomaly mediation breaking 
mechanisms in supersymmetry. These constraints include the usual LEP, B-factories, Tevatron and LHC searches, but also 
precision constraints of cosmological origin, namely the WMAP limits on the relic density of cold dark matter. We have discussed 
the standard cosmological approach and also alternative cosmological scenarios which do not change the cosmological observations
but which can affect strongly the constraints on the parameter space of these supersymmetric models based on the relic 
abundance of dark matter. We therefore show how the dark matter constraints can be weakened in order to avoid strong model 
dependent assumptions in the choice of the cosmological model. Based on different benchmark points for AMSB models, we 
performed a detailed analysis of the constraints imposed by particle data and cosmology (both standard and alternative) and 
finally we gave the typical mass spectra and decay modes relevant for the LHC searches. The main lesson that can be learnt in 
such an exercice is that usual bounds on the parameter space of these models are too restrictive and bear a strong hidden 
cosmological model dependence in the assumption of the standard cosmological scenario. Concerning the LHC searches, 
points which are excluded in the standard analysis but permitted in this more general approach, may be quite 
relevant for testing not only the particle theory models themselves but also alternative cosmological scenarios at the LHC, as in 
many cases relatively low mass supersymmetric particles are allowed with a peculiar spectrum where the lightest neutralino and chargino are very close in mass. 

\subsection*{Acknowledgements}
We would like to thank Nazila Mahmoudi for useful discussions and comments on the manuscript.


\end{document}